\documentclass[12pt,a4paper,bold]{thesis}
\usepackage{geometry}
\usepackage{amsmath}
\usepackage{amsthm}
\usepackage{amssymb}
\usepackage{setspace}
\usepackage{lscape}
\usepackage[round]{natbib}
\usepackage{graphicx}
\usepackage{epstopdf}
\usepackage{subfloat}
\usepackage[toc, page]{appendix}
\usepackage[caption = false]{subfig}
\usepackage{wallpaper}
\usepackage{color}
\theoremstyle{thm}

\theoremstyle{definition}

\providecommand{\appendixname}{Appendices}
\newcommand{\head}[1]{\newpage
\vspace{3em}
\begin{center}
\LARGE{\MakeUppercase{\textbf{#1}}}
\end{center}
\vspace{3em}
\addcontentsline{toc}{chapter}{#1}
}
\renewcommand{\listfigurename}{}

\newcommand{\thesistitle}{Power spectrum to probe turbulence in Interstellar medium using 21cm line}
\newcommand{\studentname}{Meera Nandakumar}
\newcommand{\studentrollno}{11041}
\newcommand{\advisorname}{Dr. Prasun Dutta and Dr. Rajib Saha}

\newcommand{\subject}{Physics}
\newcommand{\thesisdate}{April 2016}

\def\AA{$\AA$}

\def\maketitle{
\begin{titlepage}
\begin{center}
\begin{doublespace}
\textbf{\MakeUppercase{\LARGE{\thesistitle}}} \\
\ \\
\normalsize{\textbf{A REPORT}} \\
\normalsize{\textit{submitted in partial fulfillment of the requirements}} \\
\normalsize{\textit{for the award of the dual degree of}} \\
\ \\
\ \\
\large{\textbf{Bachelor of Science-Master of Science}} \\
\normalsize{\textit{in}} \\
\large{\textbf{\MakeUppercase{\subject}}} \\
\normalsize{\textit{by}} \\
\large{\textbf{\MakeUppercase{\studentname}}} \\
\normalsize{\textbf{(\studentrollno)}} \\
\end{doublespace}
\vfill
\centerline{\includegraphics[scale=0.20]{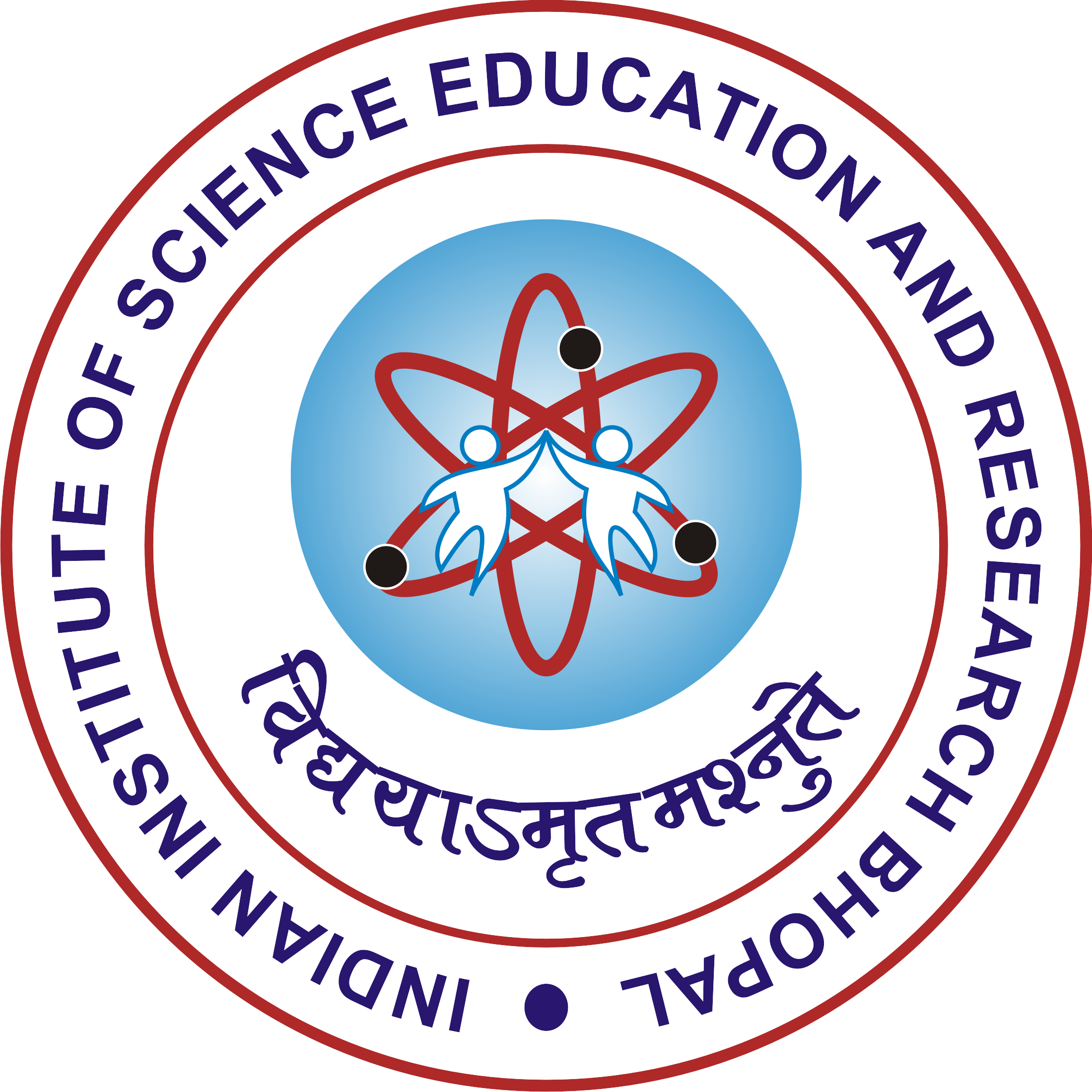}}
\textbf{DEPARTMENT OF \MakeUppercase{\subject} \\ 
INDIAN INSTITUTE OF SCIENCE EDUCATION AND RESEARCH BHOPAL\\ 
BHOPAL - 462066} \\ 
\textbf{\thesisdate}
\end{center}
\end{titlepage}
}

\begin{document}
\maketitle

\pagenumbering{roman}

\thispagestyle{empty}
\ThisULCornerWallPaper{1}{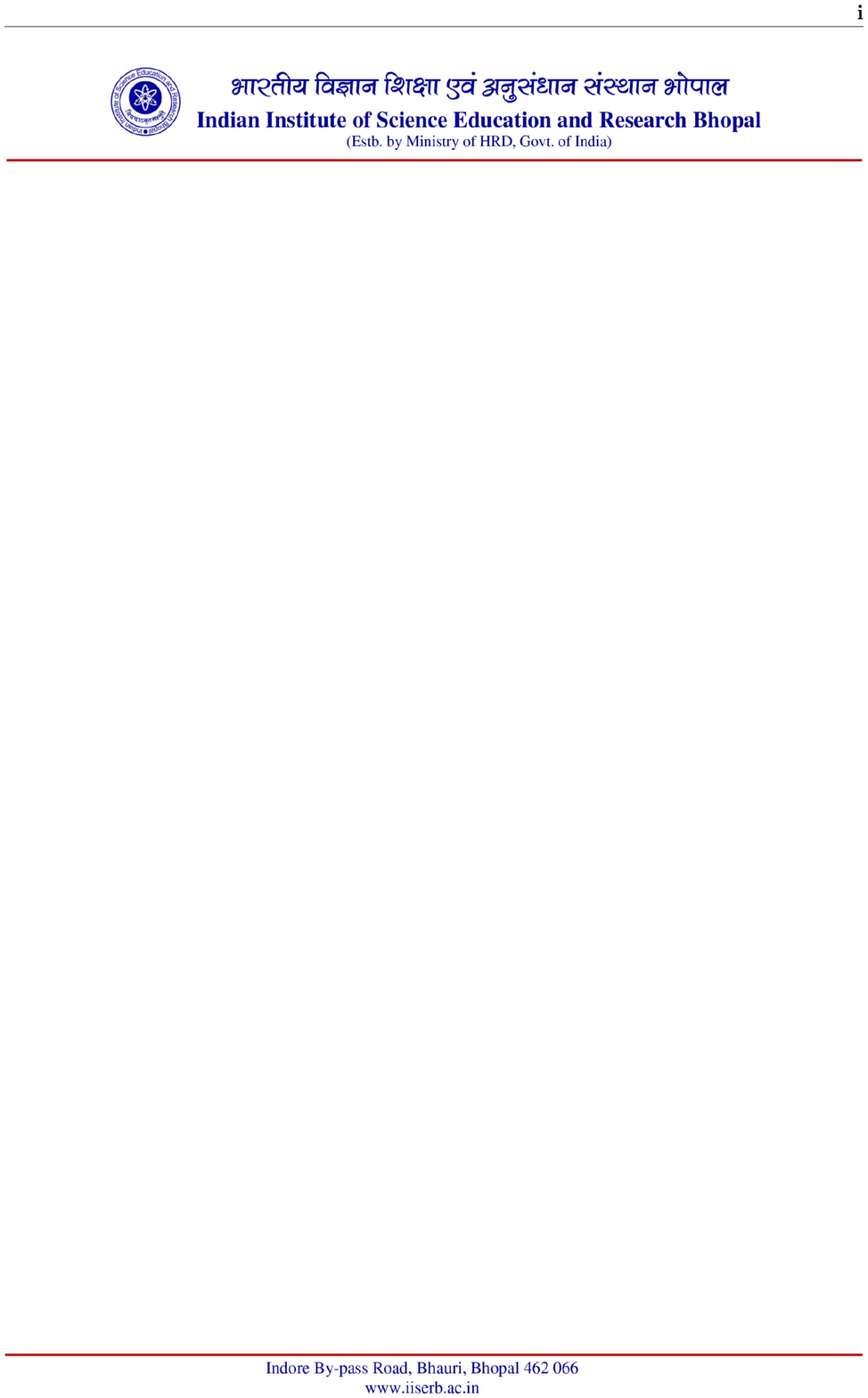}
{\color{white}mmmmmm}
\vspace{.5cm}
\begin{center}{\LARGE {\bf CERTIFICATE}}
\end{center}
\vspace{.5cm}
This is to certify that {\bf \studentname}, BS-MS (\subject), has worked on the project entitled {\bf `\thesistitle'} under my supervision and guidance. The content of this report is original and has not been submitted elsewhere for the award of any academic or professional degree.

\vspace{10em}

\textbf{\thesisdate \hfill \advisorname \\ IISER Bhopal}

\vfill

\begin{center}
\begin{tabular}{ccc}
\textbf{Committee Member} & \textbf{Signature} & \textbf{Date} \\
\\
\rule{15em}{0.4pt} & \rule{10em}{0.4pt} & \rule{6em}{0.4pt} \\
\\
\rule{15em}{0.4pt} & \rule{10em}{0.4pt} & \rule{6em}{0.4pt} \\
\\
\rule{15em}{0.4pt} & \rule{10em}{0.4pt} & \rule{6em}{0.4pt} \\
\end{tabular}
\end{center}

\head{Academic Integrity and Copyright Disclaimer}

I hereby declare that this project is my own work and, to the best of my knowledge, it contains no materials previously published or written by another
person, or substantial proportions of material which have been accepted for the award of any other degree or diploma at IISER Bhopal or any other educational
institution, except where due acknowledgement is made in the document. \\

I certify that all copyrighted material incorporated into this document is in compliance with the Indian Copyright Act (1957) and that I have received written permission from the copyright owners for my use of their work, which is beyond the scope of the law. I agree to indemnify and save harmless IISER Bhopal from any and all claims that may be asserted or that may arise from any copyright violation.

\vfill

\textbf{\thesisdate \hfill \studentname \\ IISER Bhopal}

\head{Acknowledgement}

When I was reading this thesis report for corrections, I really felt that I have done something important.  I would like to utilize this section to let others know about the persons who are responsible for getting this realization in me.
Frankly saying, if I would not be working  with Prasunda, I seriously doubt whether I may get the same feeling. The care, support, guidance, patience which he has shown is that much. When I came to know that he is  leaving IISER, I was quiet worried, whether I could continue working with him. Fortunately, I was allowed to  and I could finish the project. I am really indebted to Snigdha Mam  and Rajib Sir in this regard. 

When I thought of writing acknowledgement, one thing which suddenly came to my mind is mentioning about 14 'young' physicists: Anarto, Ganesh, Geo, Kapil, Kunal, Namitha, Piyush, Sachin, Ritwika, Shany, Shibu, Shikha, Varun and Vipin. I really miss each and every moments we have spent together for last three years. I remember the hard time we 15 people went through in courses and lab hours we have spent together in ITI. We shared our difficulties and joy and learnt Physics together.

Acknowledging my parents and my little sister Meenu, who are the  basis of my existence, is kind of awkward. While at IISER, I lossed good five  years of my life  with them, which  matter to me a lot.  At the same time, now I will really miss the same five years at IISER. It is always hard for me to accept that the days which I can spend with my friends, Archana, Jinu, Muthu, Vipin and my brother Nikhil, who had influenced me a lot, is getting reduced.  

Finally, I would like to thank the Physics department of IISER Bhopal and the whole IISER family, that has made these five years of life memorable.

\vspace{4em}

\begin{flushright}
    {\bf \studentname}
\end{flushright}

\head{Abstract}
\small
Turbulent dynamics  generate random fluctuations in density and velocity in the Interstellar Medium (ISM)  of  spiral and dwarf galaxies. Observationally, H~{\sc i} 21-cm radiation provides a good probe of these stochastic processes and helps us to know more about their nature and generating mechanism. Structure function, auto-correlation function, power spectrum are some of statistical estimators of these fluctuations which have been used in literature. Earlier studies like Dutta et al. (2013) has found that  the power spectrum of the column density fluctuations follow power law at scales at large as $\sim 10$ Kpc. However, generating mechanisms of these structures are not yet understood.  In this thesis  we are interested in estimating the velocity power spectrum at these scales to understand the energy associated with these structures and the type of forcing of the turbulence. Estimating power spectrum from observations can be either by using the directly measured quantity visibility or by reconstructing the sky brightness distribution (or image) from the visibilities. By using simulated observations of a model galaxy with known power spectrum we  find that the visibility based estimator  reproduce the true power spectrum. The image based  estimator has a scale dependent bias that  is  highly correlated to the incompleteness in the  baseline coverage of the interferometer. Interestingly, we found that the noise in the reconstructed image has zero mean and hence the locally averaged quantities estimated from the reconstructed image are unbiased. We implement the visibility moment estimator as discussed in Dutta et al. (2015)  and estimate the line of sight velocity fluctuation of the spiral galaxy NGC~6946.  Obtained velocity power spectrum obeys a power law with slope $−3.1 \pm 0.4 $ and an amplitude of $10$ km sec$^{-1}$  at scales about $4$ Kpc. Our result strongly indicate presence of large scale turbulence in the spiral galaxies and the observed slope favour slightly towards compressive forcing. We conclude that we would need to have higher signal to noise H~{\sc i} observations to definitively comment on the generating mechanisms of the large scale turbulence.


\newcommand{\sline}[2]{ {{\bf #1}} & & & &  {#2} \\}
\newcommand{\sskip}{& & & & \\}
\def\HI{H~{\sc i}\, }
\def\V{{\mathcal{V}}}
\def\P{{\mathcal{P}}}
\def\A{\mbox{\boldmath$A$}}
\def\U{\mbox{$\vec{U}$}}
\def\n{\mbox{\boldmath$n$}}
\def\k{\mbox{\boldmath$k$}}
\def\d{\mbox{\boldmath$d$}}
\def\vr{\mbox{\boldmath$r$}}
\def\vx{\mbox{\boldmath$x$}}
\def\vt{\mbox{$\vec{\theta}$}}
\def\Nt{{\bf N_{2}}}
\def\S{\mathcal{S}}
\def\N{\mathcal{N}}
\def\Wt{\tilde{W}}

\def\E{\mathcal{P}}
\def\B{\mathcal{B}}
\newcommand{\wt}{\widetilde}

\begin{singlespacing}
\vspace{-3.5cm}
\head{List of Symbols or Abbreviations}

\vspace{-1.cm}
\begin{center}
{\bf \Large{  Abbreviations}}
\end{center}
\begin{center}
\begin{tabular}{lcccl}
\sline{\large {\bf Acronym  }}{\large {\bf Full form }} 
\hline \hline
\sskip
\sline{H~{\sc i}\, }{Neutral hydrogen}
\sskip
\sline{ISM}{Interstellar Medium}
\sskip
\sline{SFR}{Star Formation Rate}
\sskip
\sline{NGC}{New General Catalogue}
\sskip
\sline{AIPS}{Astrophysical Image Processing System}
\sskip
\sline{VLA}{Very Large Array}
\sskip
\sline{GMRT}{Gaint Meterwave Radio Telescope}
\sskip
\sline{WSRT}{Westerbork Synthesis Radio Telescope}
\sskip
\sline{THINGS}{The HI Nearby Galaxy Survey}
\hline
\end{tabular}
\end{center}
\pagebreak

\begin{center}
\begin{tabular}{lcccl}
\sline{\large { \bf Symbols  }}{\large {\bf Definitions }} 
\hline \hline
\sskip
\sline{$\U$}{Baseline in kilo wavelength}
\sskip
\sline{$\nu$}{Observing frequency in MHz}
\sskip
\sline{$\vt$}{Angle in the sky measured}
\sline{}{from the centre of the galaxy}
\sskip
\sline{$\vec{r}$}{Radial vector from galactic center}
\sskip
\sline{$\V(\U, \nu)$ or $\V(\U)$}{Visibility}
\sskip
\sline{$I(\vt, \nu)$}{Specific Intensity}

\sskip
\sline{$W(\vt)$}{Window function}
\sskip
\sline{$\N(\U, \nu)$}{Noise at baseline $\U$ and frequency $\nu$}
\sskip
\sline{$P_{A}(\U)$}{Power spectrum of A}
\sskip
\sline{$M_0(\vt)$}{Zeroth moment of intensity}
\sskip
\sline{$M_1(\vt)$}{First moment of intensity}
\sskip
\sline{$V_0(\U)$}{Zeroth moment of visibility}
\sskip
\sline{$V_1(\vt)$}{First moment of visibility}
\sskip
\sline{$v_z(\vec{r})$}{Line of sight component of velocity}
\sskip
\sline{$v^T_z(\vec{r})$}{Line of sight component of turbulent velocity}
\sskip
\sline{$v^{\Omega}_z(\vec{r})$}{Line of sight component of}
\sline{}{systematic rotational velocity}
\hline
\end{tabular}
\end{center}
\end{singlespacing}



\thispagestyle{empty}

\addcontentsline{toc}{chapter}{List of Figures}
\listoffigures
\addcontentsline{toc}{chapter}{List of Figures}
\listoftables
\tableofcontents

\thispagestyle{empty}
\chapter{Introduction} \label{ch: introduction}
\pagenumbering{arabic}
Inter Stellar Medium (ISM), the medium in between the stars in our Galaxy and other spiral or dwarf galaxies, is composed of mainly gas, dust and charged particles. ISM play an important role in evolution of the spiral galaxies \citep{2011piim.book.....D}. In one end it is the seed for the star formation and then when stars die their material changes the composition of the ISM, triggering its evolution \citep{2004Ap&SS.292..193B}. Earlier observations, like by \citet{1974ApJ...193L.121J}, of the absorption lines in the stellar spectra suggested the existence of gas in between the stars. These were later called as the ISM. In depth studies of the medium followed and branched into other techniques than just optical astronomy: x-ray observations that probes the hot ionized medium, observation of molecular lines in radio and millimeter waves, low radio frequency observations of neutral hydrogen are some examples. It is understood that ISM is not a smooth passive medium, rather it has intricate structures and interesting dynamics. Theoretical understanding of these structure and dynamics indicated a turbulent medium that influence the star formation in the ISM \citep{2004Ap&SS.292..193B,2004RvMP...76..125M}) and its chemical and compositional evolution.

Systematic study of ISM had begun perhaps in 1951, when \citet{1951ApJ...114..165V}  proposed a theory of ISM, where he assumed that the differential galactic rotation stirs the entire ISM at the  large scales. It  results in  supersonic turbulence, the energy cascades down to small scales and dissipates  by viscosity.  Detailed analytical and numerical studies has addressed generation of ISM turbulence and its effects in various length scales. It is well understood that though broadly we call it ISM turbulence,  the physics of the process is different and leads to different effects at different length scales in the ISM \citep{2004Ap&SS.289..479D}. A detailed review on the progress on study of ISM turbulence and their observation can be found in \citet{2004ARA&A..42..211E}.

Turbulence generate scale invariant fluctuations in the density and velocity of gases in ISM \citep{2009ApJ...692..364F}. Since this fluctuations are intermittent and random, the information of the physical process lie in the  the statistical nature of this fluctuations. Observationally, the observation of the properties of the specific intensity of radiation  gives us informations about the dynamical variables of the process. In this thesis we address how we can probe the ISM turbulence at the largest scale, that is at the scales compared to the extent of the ISM in the galaxy. Here we discuss the backgrounds of the observational procedure and give a brief description about the  ISM of external spiral galaxies. 

\section{Statistical probes to ISM turbulence}
As turbulence being a stochastic process, the observational signatures are through various scale dependence statistical estimators of the density and velocity fluctuations of the ISM. Kolmogorov in 1941 formulated a theory of turbulence in incompressible fluid. He assumed that the system is stirred at the largest scale by some external force and the energy then percolates to the smaller and smaller scales till it reaches the scale of dissipation. In between the driving scale and the dissipation scale lies the inertial range, where rate of energy input at a scale equals the rate of energy that goes out from that scale. This results in a system with no prefered scales in between the driving and dissipation scale\footnote{See \citet{1995tlan.book.....F} for a detailed description of Kolmogorov Theory.}.  Later when the nonlinear dynamics of the compressible systems were studied, it is realized that with the velocity fluctuations the density fluctuations of the compressible fluid, that is gas, would assume scale free nature in the inertial range for a detailed description of Kolmogorov theory. Hence, to observationally probe turbulence and find out the inertial range, driving and dissipation scales and mechanisms, it is useful to look at the statistics of the density and velocity fluctuations at different length scales \citep{1972fct..book.....T}. Here are a few statistical descriptor of the same. 

The basic statistical tools are mean, median, skewness and kurtosis, which all are one point statistics of the data. To probe turbulence, it is rather important to look at the two point and higher order statistics. 
Structure function of order p for a stochastic observable A is defined as,
\begin{equation}
S_P(|\delta\vec{r}|) = \left< |A(\vec{r})-A(\vec{r}+\delta\vec{r}) |^P \right>,
\end{equation}
and theoretically if evaluated to all orders it contains all the information of the stochastic fields. 
Statistical nature of Gaussian random fluctuations  can be completely specified by its mean and either the auto-correlation function or power spectrum. 
The auto-correlation function $\xi_A(|\delta\vec{r}|)$ of any homogeneous and isotropic scalar field $A$ is defined as,
\begin{equation} 
\xi_A(|\delta\vec{r}|) =  \left< |A(\vec{r})A(\vec{r}+\delta\vec{r}) | \right>.  
\end{equation} 
Power spectrum is the Fourier transform of the auto-correlation function,
\begin{equation} 
P_A(k) =  \int d\vec{r} e^{i\vec{k}.\vec{r}} \xi_A(|\delta\vec{r}|)  
\end{equation} 
where $k=|\vec{k}|$. 
Angular brackets in the above expressions stands for  ensemble average. For astronomical observations, we are always limited to have a single realization of the sky, assumptions like statistical homogeneity and statistical isotropy are evoked to perform the above averaging. 

\section{21 cm radiation from external spiral galaxies}
Spectral lines provide the window to measure the ISM structures and dynamics. Spectral signature of the elements not only tell us the abundance and density fluctuations of the element in the ISM, but the width and the Doppler shift of the line provides the information of its dynamics. Key is to choose an element with easily observable spectral signature. Examples of spectral lines from ISM that has relatively higher strength are    Ca$^+$, Na, CO etc \citep{1970ApJ...161L..43W, 1974ARA&A..12..279Z, 1974ApJ...189..441G}. However, these elements constitute a very small part of the ISM and  distributed unevenly over the galactic disk. Since 70\% of the ISM is neutral hydrogen atom (\HI), observing it would trace the ISM more completely. The  hyperfine structure transition line from \HI (frequency 1420 MHz), other wise known as 21 cm line, provides the nicest probe to ISM structure and dynamics. 

Specific intensity of the radiation of the 21-cm line emission is given by \citep{2011piim.book.....D},
\begin{equation}
\label{eq:I_theta}
I(\vec{\theta},\nu) = I_0 \int dz \hspace{2.5pt} n_{HI}(\vec{r}) \hspace{2.5pt} \phi(\nu)
\end{equation}
where z is the line of sight direction, $\vec{r} = (x,y,z) = (\vec{R},z)$. Here $\vec{R}$ is in the plane of the sky.  The angular sepeartion of a direction in the field of view with respect to the field center is  $\vec{\theta}=\frac{\vec{R}}{D}$. $D$ is the distance between the observer and the galaxy. Note that the galaxy constitutes a very small angle in the sky, hence we can assume the sky to be flat. $I_{0}=\frac{3}{16\pi}h\nu_0 A_{21}$, where $A_{21}$ is the Einstein coefficient for 21-cm radiation. $n_{HI}(r)$ is the number density of hydrogen atom and $ \phi(\nu) $ is the line shape function. As in the different direction in the sky the velocity of the \HI gas element emitting 21-cm radiation can be different, the observed frequency of the emission is Doppler shifted. Hence, the Doppler shift can be used to write equation 1.4 in terms of line of sight velocity of the gas as 
\begin{equation}
\label{eq:I_theta}
I(\vec{\theta},v) = I_0 \int dz \hspace{2.5pt} n_{HI}(\vec{r}) \hspace{2.5pt} \phi(v).
\end{equation}
The line shape function $\phi(v)$ is  defined as,
\begin{equation}
 \phi(v)= \phi_0 \hspace{2.5pt}exp{\left[- \frac{[v - v_z(\vec{r})]^2}{2\sigma^2}  \right]}
\end{equation}
where $v_z(\vec{r})$ is the line of sight component of velocity of \HI gas and $\sigma = \sqrt{\frac{k_B T}{m_{HI}}}$ is the thermal velocity dispersion. The line shape function  is  normalized as,
\begin{equation}
\label{eq:phi}
\int dv \hspace{2.5pt} \phi(v) = 1.
\end{equation}

Moments of the specific intensity provides information about the density and velocity structures of the gas \citep{2009PhT....62e..56B}. Here we describe only first two moments. The zeroth moment of the intensity  is defined as,
\begin{equation}
M_0(\vec{\theta})= \int dv \hspace{2.5pt} I(\vec{\theta},v) 
\end{equation}
where the velocity integral is over the entire spectral range of \HI emission. The column density of the \HI gas along line of sight direction is given by $N_{HI}(\vec{\theta}) = \int dz \hspace{2.5pt} n_{HI}(\vec{r})$.   Clearly, the zeroth moment gives  the column density of the \HI gas,
\begin{equation}
M_0(\vec{\theta}) = I_0 N_{HI}( \vec{\theta}).
\end{equation}     

The first moment of $I(\vec{\theta},\nu)$ is defined as,
\begin{equation}
M_1(\vec{\theta}) = \frac{ \int dz \hspace{2.5pt} v  \hspace{2.5pt} I(\vec{\theta},v) 
 }{ M_0(\vec{\theta})} 
\end{equation}
$M_1(\vec{\theta})$ essentially gives the density weighted line of sight component of velocity of \HI gas. This indicates that using the observed specific intensity of the 21-cm radation it would be possible to estimate the statistical properties of the density and velocity of the \HI.

In this work we would like to probe the ISM of the external spiral galaxies. ISM in the spiral galaxies have morphology of a disk with the number density profile roughly exponential along the radial direction, whereas roughly it follow a Gaussian function along the direction perpendicular to the disk. The characteristic scale along the radial direction is called the scale length and that in the vertical direction is termed as the scale height. The ratio of the scale height to scale length in a typical spiral galaxy is $1:10$ or even higher suggesting that the disk is thin \citep{2004MNRAS.352..768K}. The number density of \HI in the disk, apart from this smooth variation, has statistical fluctuations in scales at 10 kpc to sub parsecs \citep{1983A&A...122..282C, 2013NewA...19...89D}. The major dynamical feature of the disk is its differential rotation with the rotation axis roughly aligned perpendicular to the disk. Added to this systematic rotation, there would be random motion of the ISM owing to the turbulent dynamics. For a given galaxy, the axis of systematic rotation can be oriented at a random direction in the sky. This orientation of the disk is usually quantified by two angles, the inclination angle and the position angle. Most of the galaxy disks however have warps, that is the position angle and the inclination angles changes with the distance from the galactic centre. The tangential rotation velocity added with the random velocity from turbulence, the inclination and position angles, gives rise to a particular pattern in the line of sight velocity. This is what can be estimated from the moment one map of the specific intensity.

\section{Radio Interferometric observation of Neutral Hydrogen}
In the earlier days, most of the understandings on ISM were obtained mainly from the Milky way galaxy using single dish radio telescopes (see for example \citet{1966AuJPA...1....3M}). Single dish radio telescopes have poor angular resolution so that it is inadequate to study the large scale structures of external galaxies. Meanwhile, radio interferometers  are composed of many different array elements, called antenna or tiles, which effectively act as single dish radio telescope with high resolution. So, measurement of the \HI intensity at a large range of length scales, requires radio interferometric arrays \citep{1983A&A...122..282C}. VLA(Very Large Array), GMRT(Gaint Meterwave Radio Telescope), WSRT(Westerbok Synthesis Radio Telescope) are some of the important radio interferometers in the world.  In an interferometer, every pair of these antenna measures a quantity called visibility  $\mathcal{V}(\vec{U},\nu)$, which can be approximated as the Fourier transform of the specific intensity $I(\vec{\theta},\nu)$. Visibilities are measured discretely at the baseline vector $\vec{U}$ given by the instantaneous projected separation of the antenna pair along the plane of the sky in units of observing wavelength. Hence, the interferometers effectively sample the Fourier transform of $I(\vec{\theta},\nu)$, i.e $\tilde{I}(\vec{U},\nu)$ at a set of discrete points in $\vec{U}$ given by the array configuration of the antenna, declination of the source and the integration time of observation. The measured visibilities can be written as
\begin{equation}
\mathcal{V}(\vec{U},\nu)=\tilde{I}(\vec{U},\nu)S(\vec{U})
\end{equation}   
where, $S(\vec{U})$ is the sampling function. If the total number of sampling in the baseline space is  $N_{b}$, the sampling function would be
\begin{equation}
S(\vec{U})=\sum^{N_{b}}_{i=1}\delta(\vec{U}-\vec{U_{i}}).
\end{equation}
Inverse Fourier transform of the measured visibility is called the dirty image:
\begin{equation}
I_{D}(\vec{\theta},\nu)=I(\vec{\theta},\nu) \otimes B(\vec{\theta}).
\label{eq:psf}
\end{equation} 
 Here $B(\vec{\theta})$ is the Inverse Fourier transform of the sampling function and essentially the  Point Spread Function (or  beam) of the interferometer.
 Since the sampling function can be quite discrete, the interferometer  beam is non trivial and the dirty image can not be used as an estimate of the sky image. It is necessary to deconvolve the interferometric beam from the dirty image. Strictly speaking, in most of the cases when the sampling in the baseline space is sparse, specially with extended emission, the deconvolution does not give unique results. Nevertheless, different deconvolution schemes are used, CLEAN, MEM are to name a few. Chapter 2 describes CLEAN algorithm \citep{2008AJ....136.2897R}, which we are using in our work. Finally, we note here that the observed frequency shift can be directly transferred to the velocity of the gas that is emitting the \HI radiation as discussed in the previous section.

\section{Aim of this thesis}
Column density power spectra of Milky way galaxy was found to obey power law at length scales ranging sub parsacs to $200$ pc, suggesting the existence of turbulence there in. It was argued that the turbulence is mostly generated as a results of the supernovae shocks stirring the ISM. 
Recent estimation of \HI column density power spectrum from external dwarf and spiral galaxies by \citet{2006MNRAS.372L..33B}, \citet{2008MNRAS.384L..34D,2009MNRAS.398..887D,2013NewA...19...89D} found that they obey power law ranging from 1  kpc to 10 kpc. It is unlikely that  these large scale  fluctuations are a result of turbulence driven by supernovae. Numerical studies by \citet{2009ApJ...692..364F} demonstrated the nature of solenoidal vs compressive forcing in compressible fluid turbulence.  An observational probe in the velocity fluctuation will clearly suggest more on the character of the energy input and the generating mechanism of these large scale structures. We plan to investigate in this line here.

There are several quantifier of the scale dependence of the velocity fluctuations of the ISM of our galaxies, like Velocity channel anaslyis(VCA)\footnote{VCA: \citet{2000ApJ...537..720L}}, Velocity Coordinate Spectrum (VCS)\footnote{VCA: \citet{2008arXiv0811.0845C}}, statistics of the centroid of velocities\footnote{\citet{2007MNRAS.381.1733E}} etc. Among these VCA is most efficient technique \citep{2001ApJ...551L..53S, 2015ApJ...810...33C}, and
has been used extensively to probe the small scale velocity structures of our galaxy. Recently, \citet{2016MNRAS.456L.117D} demonstrated that VCA has limitations when applied to external galaxies. 

There are two distinct classes of statistical estimators. Some of these are based
on the directly measured quantity from the interferometers, the visibilities, others rely on the
reproduction of the sky brightness distribution from the interferometric data. It has been shown that
in some cases these two different techniques results in conflicting estimates of the statistical quantities, like the power spectrum.
Though visibility based estimators are more direct, the image based estimators can estimate the
power spectrum of parts of the field of view of the telescopes. Later is essential at some particular
cases, like to correlate star formation with ISM turbulence, variation of MHD turbulence in arm
and inter arm regions of the spiral galaxies etc. In fact the ISM velocity structure estimators that we have outlined above all rely on the reconstructed image. 

In this work we first quantify the efficacy of the 
image and visibility based estimators of the power spectrum using numerically simulated observations. We use a model
\HI observation from an external face on spiral galaxy for this purpose. We investigate the reason
for the possible deviation from the true power spectrum. Our investigation suggested that the visibility based estimators are unbiased and more desired. Next we implement a visibility based estimator for the 
 velocity power spectrum of spiral galaxies and draw scientific conclusions from the measurements. Remaining part of the report is divided as follow. Chapter 2 discuss about the efficacy of two power spectrum estimators. The velocity power spectrum estimator we have used is discussed in Chapter 3, we also note our result there.  At the end, we conclude the report in Chapter 4 with our findings and its impact on ISM physics.

\chapter{On the merit of the power spectrum estimators}
\label{ch:introduction}
Power spectrum quantifies  the scale dependence of the  random fluctuations. In this chapter we discuss  the image and visibility based power spectrum estimator and  check the efficacy of them using numerically simulated observations. We also comment on  the reason for  the possible deviation. 

\section{Power spectrum estimation}
The moment zero map of the specific intensity, $M_0(\vec{\theta})$, gives the column density of the \HI gas. The power spectrum of the column density  is defined as
\begin{equation}
P_{HI}(\vec{U})=\delta_{D}(\vec{U}-\vec{U}') \left \langle \tilde{M}_{0}(\vec{U})^{*}\tilde{M}_{0}(\vec{U}') \right \rangle \footnote{We denote the two dimensional Dirac delta function by $\delta$ with a suffix D. Tilde '~' denotes the corresponding Fourier transforms.}
\end{equation}
The angular bracket here refer to ensemble average and has to be taken over many realisation of the sky. In practice, the  fluctuations  are often statistically isotropic, the ensemble average can be replaced with an azimuthal average. Following, we discuss two different estimators that has been used in literature to probe the power spectrum of \HI intensity fluctuations from radio interferometric observations.

\subsection{Visibility based power spectrum estimator}
Visibility based power spectrum estimator was introduced by \citet{2001JApA...22..293B}.  It has been widely used to find out the power spectrum of \HI intensity fluctuation of our Galaxy and external spiral and dwarf galaxies and also is proposed as a major tool in detecting the \HI 21 cm signal from the epoch of reionization. Here we give a brief overview. For more technical details, readers may refer to texts like \citet{2011arXiv1102.4419D}. We start with investigating the quantity
\begin{equation}
P_{V}(\vec{U}) = \delta_{D}(\vec{U}-\vec{U}') \left \langle V^{*}(\vec{U})V(\vec{U}') \right \rangle,
\end{equation}
and use the expression for  visibility  in eqn~(2)  to write above as
\begin{equation}
P_{V}(\vec{U}) = \delta_{D}(\vec{U}-\vec{U}') \left \langle \tilde{M_{0}}(\vec{U})^{*}\tilde{M_{0}}(\vec{U}') \right \rangle \left \langle S^{*}(\vec{U}) S(\vec{U'})\right \rangle.
\end{equation}
Here we have used the fact that the fluctuations in the sky and the sampling function are uncorrelated. The measured visibilities directly gives $P_{V}$.  For a particular observation the sampling function is well known and the second term can also be estimated completely. The first term in the angular bracket is essentially the \HI power spectrum defined in eqn.~(5). Since the effect of the sampling function here is just a multiplicative factor, the \HI power spectra can be directly estimated from the visibilities. This estimator is usually refered to as the visibility correlation estimator in literature. In practice, the measured visibilities also accompany measurement noise from the interferometer. However, such a noise is independent of baseline and can be trivially removed from the power spectrum estimates \citep{2008MNRAS.384L..34D}.

Visibility based power spectrum estimator works with the directly observed quantity, the visibility,  and hence is a more direct probe of the power spectrum from the data. One do not need to go through a complex deconvolution procedure here. On the other hand, as the power spectrum is calculated in visibility space, it is not straight forward to select a part of the field of  view of the interferometer and selectively estimate the power spectrum for that part. This is a major shortcoming of this estimator.

\subsection{Image based power spectrum estimator}
Image based power spectrum are also used in literature, where the sky image estimated through deconvolution of the interferometer beam from the dirty image is used. Different deconvolution algorithm are used to reconstruct the sky brightness distribution, among which CLEAN is widely used. Here we briefly describes about the CLEAN algorithm. 
In CLEAN, the sky image is assumed to be a collection of point sources. It identifies the brightest point sources in the sky from the brightest pixels in the dirty image and (partially) remove the effect of these sources in the visibilities. It them remake the dirty image and proceed in a similar manner. Thus it uses a simple iterative procedure to find the position and strength of (all) the point sources. The final deconvolved CLEAN image is these point sources convolved with a synthesized beam. The synthezised beam is considered to be the best fit  Gaussian to the primary beam of the interferometer. Residual flux, after modeling all the point sources from the dirty image, is added to the above reconstruction. We shall refer to the estimate of the sky image using clean by the quantity $M_{C}(\vec{\theta})$ and would call it the CLEAN image.

The image based power spectrum is defined as
\begin{equation}
 P_{HI}^{(I)}(\vec{U})=\delta(\vec{U}-\vec{U}') \left\langle \tilde{M}_{C}(\vec{U})\tilde{M}_{C}^*(\vec{U}') \right\rangle,
\end{equation}
where $\tilde{M}_{C}$ is the Fourier transform of the CLEAN  image. As the power spectrum is calculated from the image here, it is possible to select a particular region in the field of view and estimate power spectrum of it. 

A typical interferometer like GMRT\footnote{GMRT: Giant Meterwave Radio Telescope}, VLA\footnote{Very Large Array}, WSRT\footnote{Westerburg Synthesis Radio Telescope}, LOFAR\footnote{Low Frequency Radio Array} etc samples the visibility functions only at specific points in the baseline space. Any deconvolution procedure that try to reduce the effect of the interferometer beam from the dirty image, requires interpolation of the visibilities at the non sampled baselines  and hence introduces spurious correlation among visibilities at different baselines. This, manifests as a correlated noise $N_{C}(\vec{U})$ in $\tilde{M}_{C}(\vec{U})$, where 
\begin{equation}
\tilde{M}_{C}(\vec{U}) = \tilde{M}_{0}(\vec{U}) + N_{C}(\vec{U}).
\end{equation}
Hence, the image based estimator can be written as 
\begin{equation}
 P_{HI}^{(I)}(\vec{U}) = P_{HI}(\vec{U}) +  P_{N_{C}}(\vec{U}),
\end{equation}
$P_{N_{C}}(\vec{U})$ is the power spectrum of the correlated noise.
Though we expect that the  noise bias depends on the baseline coverage of the particular observation, an analytical estimation of it is not straight forward. Interestingly, such a bias is grossly ignored in literature where  image based estimators are used  \citep{2012ApJ...754...29Z}.

In this work, we perform a controlled test on the efficacy of the two estimators of the power spectrum of the \HI column density discussed above. We proceed as follows. We generate a sky model that has a  correlated column density fluctuation with a known power spectrum. We perform simulated observation assuming a baseline distribution of the interferometer to get the observed visibilities. We use the  IMAGE task in AIPS \footnote{NRAO-AIPS: Astrophysics Image Processing System} that uses the CLEAN algorithm to deconvolve the interferometer beam from the dirty map and make an estimate of the sky image. We estimate $P_{HI}^{(I)}(\vec{U})$ from this deconvolved image and also estimate $P_{HI}^{(V)}(\vec{U})$ directly using the simulated visibilities. We finally compare these power spectrum estimates with the power spectrum of the input model image.

\section{Simulating H~{\sc i} observations from a spiral galaxy}

\subsection{Sky Model}
We model the moment zero map of the \HI intensity (image)  in the following way
\begin{equation}
M_{0}(\vec{\theta})=W(\vec{\theta})\left[ \bar{M}_{0} + \delta M_{0}(\vec{\theta})  \right],
\label{eq:mod}
\end{equation}
where $W(\vec{\theta})$ quantifies the large scale \HI distribution in the sky and is normalized as 
\begin{equation}
\int W(\vec{\theta}) d\vec{\theta} = 1.
\end{equation}
In case of observations where the \HI emission is spread over the entire field fo view of the interferometer, the primary beam of the interferometer gives $W(\vec{\theta})$. For observations related to external galaxies, where mostly the galaxies are localized to a small part of the field of  view of the telescope, $W(\vec{\theta})$ quantifies the large scale distribution of \HI column density in the galaxy. The quantity $ \bar{M}_{0}$ is proportional to the total intensity coming form the entire field of view and can be written as 
\begin{equation}
 \bar{M}_{0} = \int M_{0}(\vec{\theta}) d\vec{\theta}.
\end{equation}
The component $\delta M_{0}(\vec{\theta})$ gives the fluctuation in the \HI column density. We assume it to have a zero mean, i.e, $\langle\delta M_{0}(\vec{\theta}) \rangle  = 0$. We shall discuss how we model the window function and the fluctuations shortly.

\subsubsection{Modeling window function}
\HI profile of spiral galaxy is dominated by a radial variation in \HI column density. However, azimuthal variations, like spiral arms, rings, are also seen. We  model the window function based on large scale structure of the face-on spiral galaxy NGC~628. As we want to keep anisotropic large scale features of the \HI distribution in  the window function, we use `shapelets' here to represent it. We decompose the moment zero map of NGC~628 taken from THINGS survey\footnote{THINGS: The \HI Nearby Galaxy Survey \citet{2008AJ....136.2563W}} data product  in terms of it's shapelet coefficients and use first few shapelets to model the window function. In interest of completeness of this report, we first give a brief description of  `shapelet' here and then discuss the criteria we use to choose the parameters of the shapelet reconstruction.

\begin{figure}[t!]

\subfloat[]{\includegraphics[scale=.32]{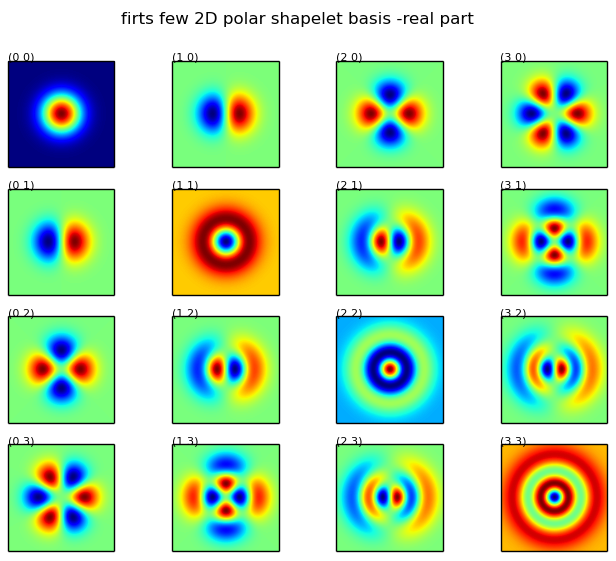}}
\subfloat[]{\includegraphics[scale=.32]{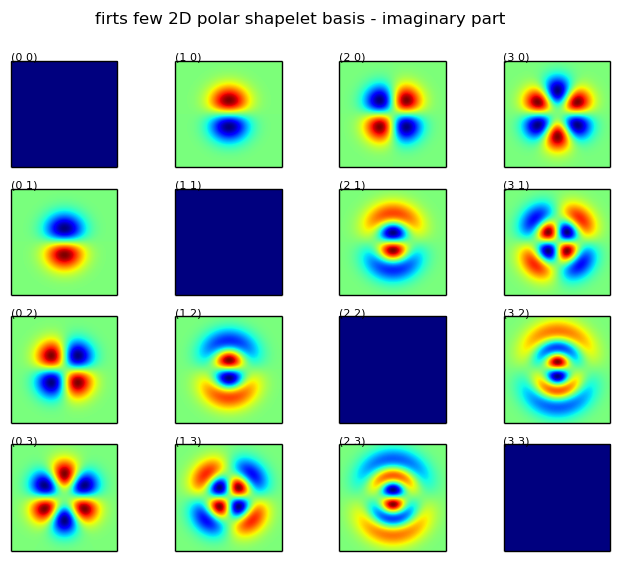}}
\caption{Fig~(a) and Fig~(b) shows the real and imaginary parts of the first few two dimensional polar shapelet basis functions respectively.}
\label{fig:polar}
\end{figure}
Shapelets are defined as a set of localized  basis functions with different shapes \citep{2003MNRAS.338...35R}, we use Gaussian weighted Hermite polynomials in polar coordinates here. These forms a complete orthonormal basis for smooth, integrable functions and hence  any well behaved 2D functions  can be decomposed into shapelets. Polar Hermite polynomials can be written using recursion formulae,
\begin{eqnarray}
\frac{l-k}{x}H_{k,l}(x)&=&lH_{k,l-1}(x)-kH_{k-1,l}(x), \ \ \ \ for\  k\neq l \\ \nonumber
H_{k,k}(x)&=&H_{k+1,k-1}(x)-x^{-1}H_{k,k-1}(x) \ \ \ \ for\  k = l. 
\end{eqnarray}
Shapelets are defined as,
\begin{equation}
S_{n_l,n_r}(r,\phi, \eta)= \sqrt{\pi n_l!n_r!}\ \eta^{-1}H_{n_l,n_r}(x/\eta)e^{-\frac{x^2}{2\eta^2}}e^{i(n_l-n_r)\phi}.
\end{equation}
Here $r$ is the radial and $\phi$ is the angular coordinate and $\eta$ corresponds to the scale of the shapelets. 
The first few order of polar shapelets are shown in the figure~\ref{fig:polar}. Any well behaved function $f(r, \theta)$ can be decomposed in terms of its shapelet coefficients $f_{n,m}$ as 
\begin{equation}
f(r,\phi)=\sum_{n=0}^{\infty}\sum_{m=-n}^{n}f_{n,m}S_{n,m}(r,\phi,\eta).
\end{equation}

In order to model the window function  using the moment zero map of the galaxy NGC~628, we need to choose the scale of the shapelets, i.e, $\eta$. We do this in the following way. Considering a given value of $\eta$, we construct the zeroth order shapelet (a Gaussian function essentially) from the moment zero map of NGC~628 and estimate the mean square difference between the moment zero map and this basic shapelet. We choose a value of $\eta$ which corresponds to the lowest mean square difference as estimated above. 

\begin{figure}[t!]

\subfloat[]{\includegraphics[scale=.25]{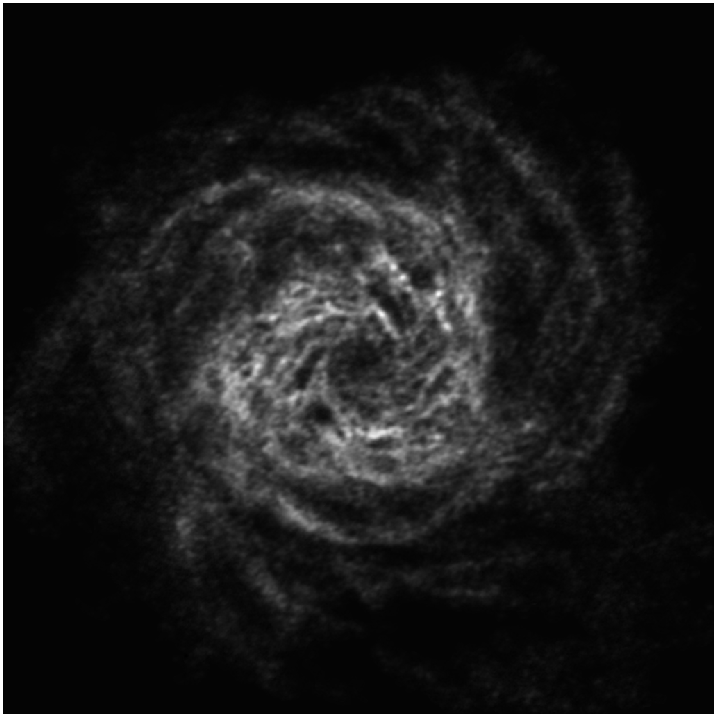}}
\hspace{10pt}
\subfloat[]{\includegraphics[scale=.51]{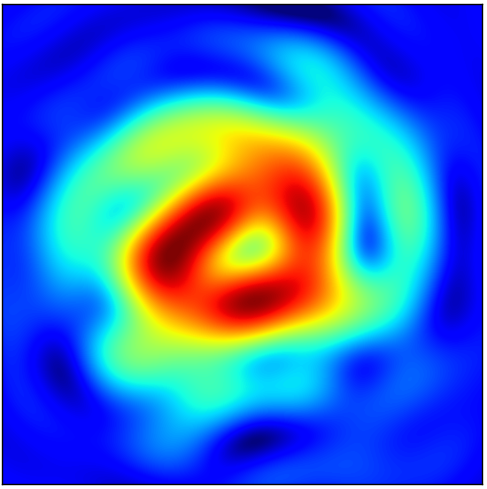}}
\caption{Fig(a) is moment zero map of the \HI emission distribution of the NGC~628 galaxy obtained from THINGS survey. Fig(b) is model of the window function that we use for the simulation.  }\label{fig:NGC628}
\end{figure}

Dutta et al. (2013) have used a visibility based power spectrum estimator to estimate the \HI intensity fluctuation power spectrum of the galaxy NGC~628  from THIGNS observations. They found at baselines $> 1\ k\lambda$, the power spectrum is well fitted by a power law. At smaller baselines, the large scale structure of the galaxy, i.e, the window function dominates.  We found that for shapelet reconstructions with shapelets higher than of order 12, the window function has an significant effect at baselines $> 1 \ k\lambda$. Hence, the maximum value of the shapelet order that we use here is 12. Using these parameters and the normalization criteria given in eqn.~(12), we construct the model window function. Figure~\ref{fig:NGC628} shows the moment zero map of the galaxy (a) NGC~628 and the model window function (b) constructed based on it.

\subsubsection{Modeling the $\bar{M_{0}}$ and $\delta M_{0}(\vec{\theta})$}
It is to be noted that the absolute amplitudes of the quantities $\bar{M_{0}}$ and $\delta M_{0}(\vec{\theta})$ are not of much interest apart from a scaling of the model image. The relative amplitudes of $\bar{M_{0}}$ and standard deviation  of $\delta M_{0}(\vec{\theta})$, i.e, $\sigma_{\delta M_{0}}$,  need to be fixed for the simulation. We call this quantity $\mathcal{R} = \bar{M_{0}}/\delta M_{0}(\vec{\theta})$. The random fluctuations, $\delta M_{0}(\vec{\theta})$, we model as zero mean  Gaussian random distribution with a power law power spectrum. Variance of these fluctuations provides an amplitude to the power spectrum, we choose it to be unity. We choose different values for the index of the power law spectrum $\alpha$ and $\mathcal{R}$, which will be discussed in next section.

\subsection{Simulated observations}
We use the model of the sky as discussed in the previous section to simulate radio interferometric observations and generate random group visibility fits files. For this purpose we need to choose a particular array configuration of the interferometer. We model our telescope based on the GMRT array configuration with same telescope latitude and the baselines scaled to half its original values \footnote{GMRT original array configurations can be seen in {\it http://gmrt.ncra.tifr.res.in/gmrt\_hpage/Users/doc/GMRT\-specs.pdf}}. The largest baseline of our model telescope is $60\ k\lambda$. 

\citet{2013NewA...19...89D} has estimated the power spectra of 18 spiral galaxies from the THINGS sample using a visibility based estimator. They found that the power spectra follow power laws at baselines larger than $\sim 1\ k\lambda$. The power law index $\alpha$ were found to lie between $-0.3$ to $-2.2$. Moreover,   9 of the 18 galaxies have $\alpha$ in between $-1.5$ to $-1.8$. We choose three sets of values of $\alpha$ for our model sky image : $[-0.5, -1.5, -2.0]$. \citet{2013MNRAS.436L..49D} found that the ratio of the fluctuating to the mean component varies between $5$ to $10$ for the six galaxies they have analysed. We consider two values of $\mathcal{R}$ here: $[5,10]$. We generate model sky images based on these parameters in a square grid of $1024^{2}$ with each grid element representing a $1.5^{''} \times 1.5^{''}$ patch in the sky. We choose the declination of the source to be $+54^{\circ}$ for our simulation. Using these parameters we perform 8 hours equivalent of simulated observation of the model sky. We do not include any measurement noise in the simulated observation. 

We use the AIPS task IMAGR to prepare the dirty image as well as a deconvolved estimate of the model sky. We use visibility and the image based power spectrum estimators with the simulated visibilities and the dirty as well as CLEAN image of the model sky respectively to estimate the power spectra. We also estimate the power spectra of the model sky directly from the sky model and use that as reference.
\section{Analysis of the simulated data}

\begin{figure}[t!]

\subfloat[]{\includegraphics[scale=.39]{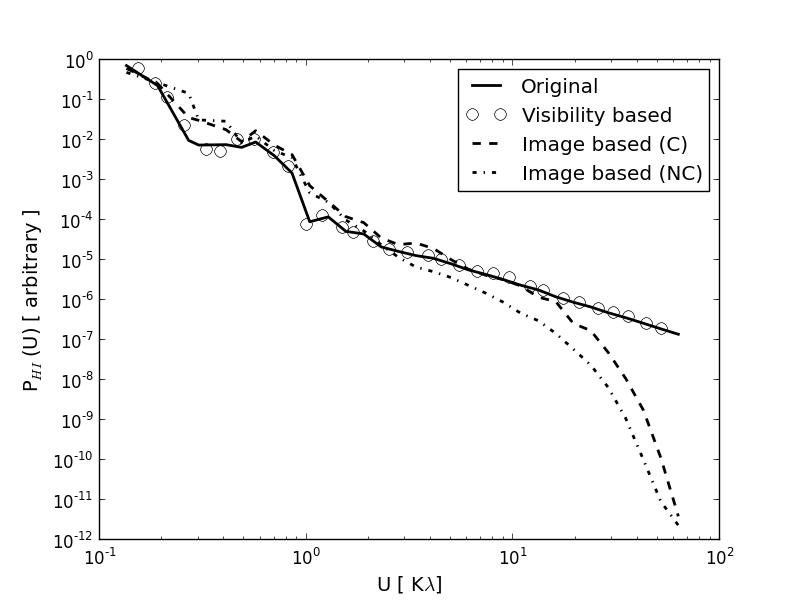}}
\subfloat[]{\includegraphics[scale=.27]{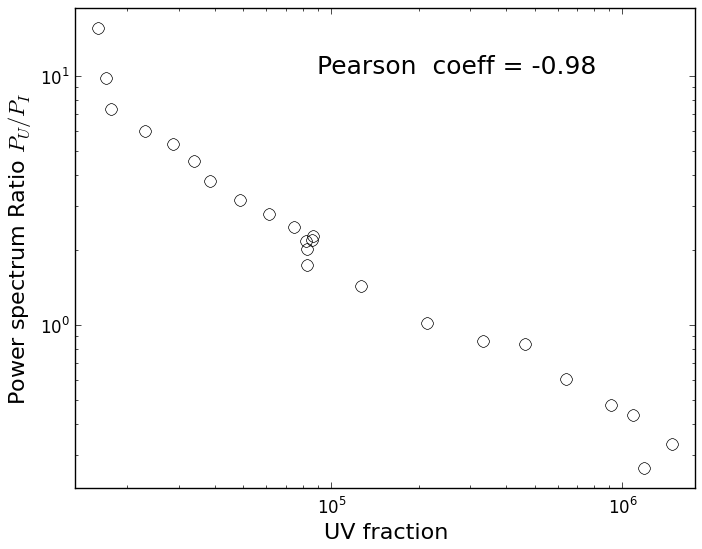}}
\caption{Fig~(a) shows the power spectrum estimations. Fig~(b) gives the correlation between the baseline fraction with the ratio of the image and visibility based power spectrum.} \label{fig:ps}
\end{figure}

\begin{landscape}
\begin{figure}[t!]
\label{fig:allps}
\includegraphics[scale=0.7]{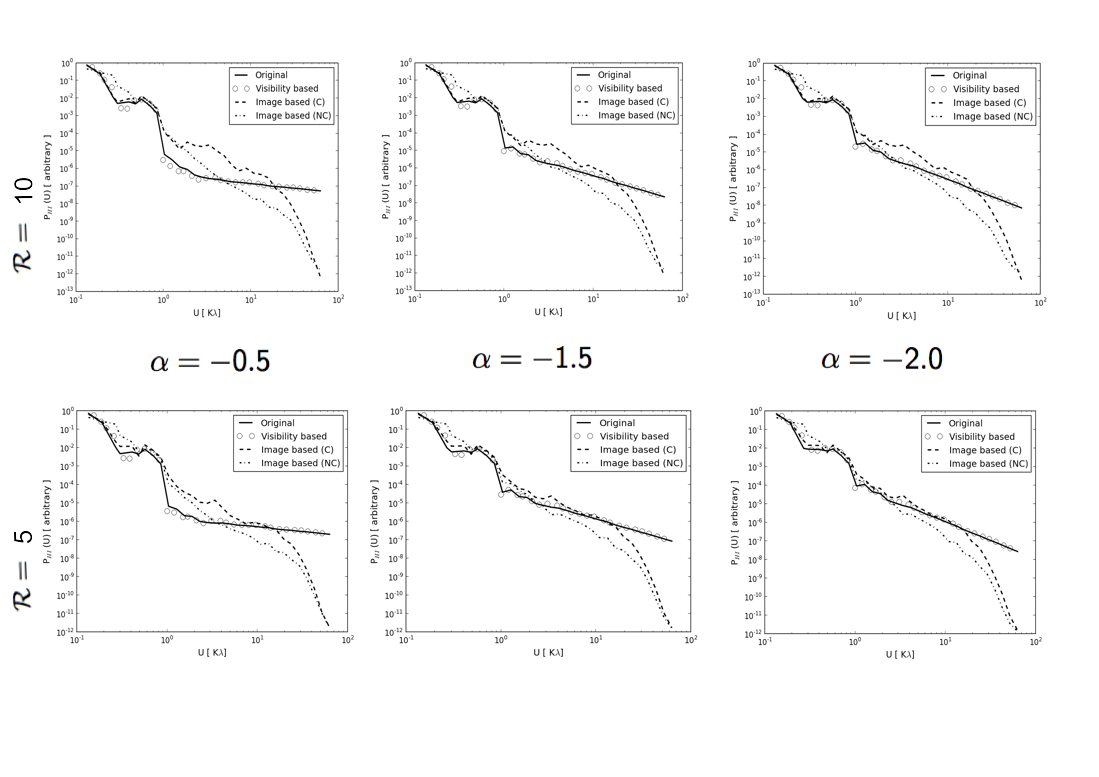}
\caption{Power spectrum comparison plots for different simulation runs.}
\end{figure}
\end{landscape}

In this section we shall compare the power spectrum estimated by the different methods. To keep things simpler, we describe the outcome for only one set of simulation in the main text with $\alpha = -1.5$ and $\mathcal{R} = 5$. All the conclusions drawn in this section can be carried forward for the rest of the models. The power spectrum plots are shown in Figure~\ref{fig:allps} at the end of this chapter.

\subsection{Power spectrum}
Figure~\ref{fig:ps}(a) summarizes the power spectrum analysis we do here, where we plot the azimuthal averaged power spectra in a log-log scale as a function of baseline $U$.  The black solid line refer to the power spectrum of the model sky and can be considered as a reference. It is clear that at baselines smaller than $1\ k\lambda$ the window function dominates, whereas at larger baselines the power spectrum assume a power law with  $\alpha = -1.5$, same that of the model image.  The open circles corresponds to the power spectrum estimated from the visibility based estimator. Clearly, it almost reproduces the reference spectrum. This demonstrates the efficacy of the visibility based power spectrum estimator to accurately  reproduce the actual intensity fluctuation power spectrum of the sky.

The dashed line and the dot-dashed line represent the power spectra calculated using the image based estimator from the dirty image and the CLEAN image respectively. Clearly, at baselines $> 1 \ k\lambda$ they deviates from the reference spectra. At these baselines, they appear to follow a power law spectra with a  steeper slope compared to the reference spectra. Note that, the extra  steepening of these power spectra at baselines $\sim 20\ k\lambda$ and higher is an effect of the convolution of the effective synthesised beam at the last stage of CLEAN. The baselines $> 20\ k\lambda$ do not carry information about the sky.

It is clear from above discussion that the image based estimator fail to reproduce the reference power spectrum. In fact, if we blindly use the image based estimator, we shall infer a higher index for the power law. 

\subsection{ UV coverage and the power spectra}
In figure~~\ref{fig:ps}(a), we see  a trend that the ratio of the image and visibility based power spectra deviates more with the increase of the baseline value. Here we investigate the possible reason for this. As mentioned in section 2, we expect the image based power spectrum to have correlated noise bias because of incomplete baseline coverage. To emphasis, while estimating the image from the observed discretely sampled visibility values, the deconvolution procedure involves interpolations of the visibility at the baselines where the visibility is not sampled. The required amount of interpolation depend on the array configuration and source declination. Hence, we would expect, in case of an array with complete sampling, the
noise bias discussed above will be absent. For example, when we estimate the power spectra from
the model image, we perform a Fast Fourier Transform  and  get the visibilities in a grid. This is
an example of complete sampling and hence we do see that the model image has the exact power
law power index as it is modeled on.
Naively, one would expect that the lack of baseline coverage would then correlate with the difference
in the power spectra estimated from the image and from the visibility. As at the large baselines
the power spectra amplitude decreases following a power law, additive difference in the two power
spectra would be less at larger baselines. A better quantifier of their true difference would be
the ratio of the two spectra which we use here. We quantify the baseline coverage by the quantity
baseline-fraction $F_{b}$ defined as
\begin{equation}
F_{b}(U)=\frac{\int_{\vec{U}}^{\vec{U}+\Delta \vec{U}}S(\vec{U})d\vec{U}}{\int_0^{\vec{U}_{max}}S(\vec{U})d\vec{U}} \frac{A_{U}}{N_{b}},
\end{equation}
where $A_{U}$ is the total area of the baseline plane.
We plot baseline fraction against the ratio of the power spectrum for the baseline range $1 - 20 \ k\lambda$ in figure~~\ref{fig:ps}(b). It is clear that at the lower  values of $F_{b}$,  the deviation of the two power spectra is more. To quantify any correlation between the ratio of the power spectra with $F_{b}$, we estimate the Pearson linear correlation coefficient, which came out to be $-0.98$ suggesting a very strong (anti) correlation. 

We come to two important conclusions here. Firstly, the visibility based power spectrum estimator accurately reproduces the power spectrum of the true sky fluctuations and can be used without any ambiguity. The image based estimator, for the observations with incomplete baseline coverage,  deviates significantly from the true spectrum and has a scale dependence bias. Use of the image based power spectrum without any definitive test would result in a biased quantification of the sky fluctuations. Secondly, we observe that the incomplete baseline  has significant effect on the image based power spectrum estimator. In fact, the lack of baseline coverage correlates highly with the deviation of the image based power spectrum from the true value. This suggest clearly that the array configuration of the interferometer has a big role to play for an image based power spectrum to be effective. 

We mention at the end that though the results presented here is based on one set of simulation with $[\alpha = -1.5, \mathcal{R} = 5.$, the same conclusions can be drawn from the other sets. We request the interested reader to have a look at Appendix-II for details.
\section{Power spectrum of THINGS images}

\begin{figure}[t!]

\subfloat[]{\includegraphics[scale=.35]{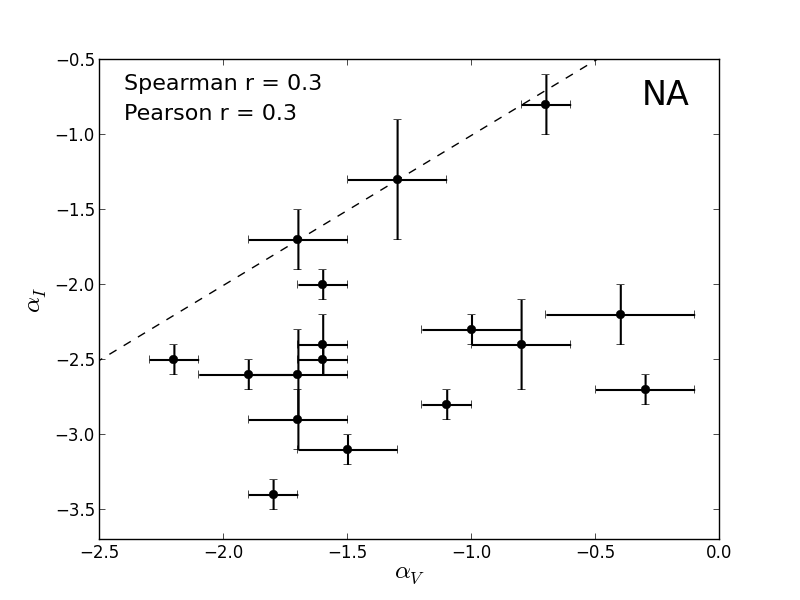}}\hspace{10pt}
\subfloat[]{\includegraphics[scale=.35]{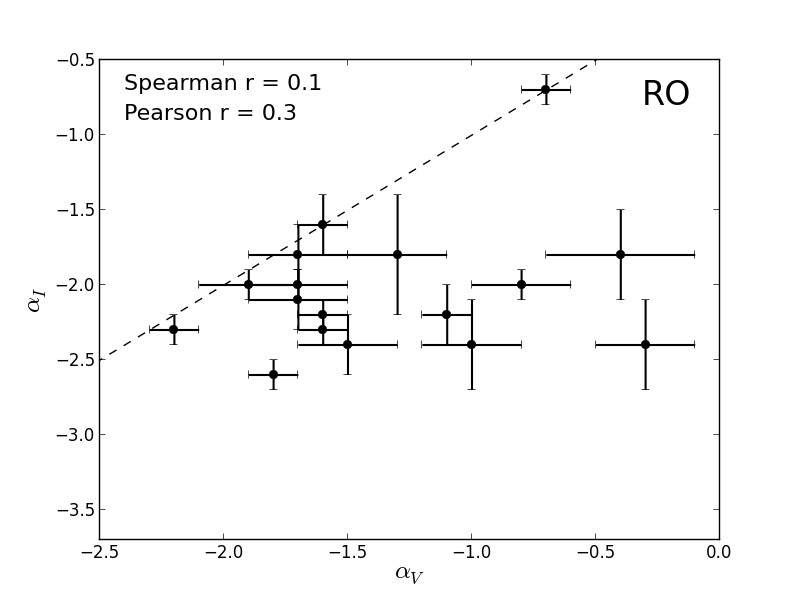}}
\caption{Figures showing the scatter plot of the power law slope for the THINGS galaxies estimated with image and visibility based methods with natural weighted [Fig~(a)] and robust weighted [Fig~(b)] images.}
\label{fig:ps1}
\end{figure}
Dutta et al. (2013) has estimated the power spectrum of the \HI intensity fluctuation of 18 spiral galaxies from THINGS sample using a visibility based estimator. They found that over a range of length scales the power spectra are well fit by power laws. As we have found from our controlled test using  simulation, that the visibility based estimator reproduce the intensity fluctuation power spectra almost exactly, we shall use the estimates reported by Dutta et al (2013) as proxy for the true \HI power spectra of these galaxies. We use the image based estimator described in section~(3.2) to estimate the power spectra of the same galaxies.  We find that for a range of length scales, for each of the galaxies, the power spectra follow power laws. We perform a regression analysis on the image estimated power spectra  to find the slope of these power laws. Note that the highest baseline that a visibility based power spectrum estimator is limited to depends on the measurement noise in visibilities, whereas ignoring the noise bias, the highest baseline that an image based power spectra probe is given by the inverse of the interferometer beam. The visibility data used by Dutta et al (2013) lack the measurement at the zero baseline. However, in addition to the visibilities, the total \HI flux of the galaxy estimated from single dish observations are also used to construct the moment zero maps in the THINGS data product. As this changes the power at the smaller length scales, the window function dominance in the power spectrum may extend to different baselines for the image and visibility based cases. Hence, we do not restrict ourself to choose the same range of baselines as in the visibility based case to find the power law fit for the image based case. We rather choose the best possible extent in the baseline to fit  power laws to the image based power spectra. The power law slope, errors, the geometric mean of the range of baselines ($U^{(G)} = \sqrt{U_{min} U_{max}}$) they are fit to for each galaxies are listed in table~(1) for both visibility and image based estimators. Note that, we represent the best fit power law index from the visibility based power law estimator as $\alpha_{V}$ whereas the best fit power law index estimated from the image based estimator is termed as $\alpha_{I}$. Further, THINGS data products provide both robust weighted (RO) and natural weighted (NA) moment zero maps. We perform the analysis described in this section for both these maps.
\begin{table}
\begin{center}
\begin{tabular}{lcc|cc|cc}
\hline
 & Visibility & based & Natural & weighted & Robust & weighted \\
Galaxy & $\alpha_{V}$  & $U^{(G)}_{V}$ & $\alpha^{(NA)}_{I}$ & $U^{(G)}_{I}$ & $\alpha^{(RO)}_{I}$ & $U^{(G)}_{I}$ \\
\hline
\hline
 & & & & & & \\
NGC628  & -1.6$\pm$0.1 & 3.2 & -2.4$\pm$0.2 & 4.0 &     -2.3$\pm$0.1  & 4.5 \\ 
NGC925  & -1.0$\pm$0.2 & 3.2 & -2.3$\pm$0.1 & 5.5 &     -2.4$\pm$0.3  & 4.4 \\ 
NGC2403 & -1.1$\pm$0.1 & 2.2 & -2.8$\pm$0.1 & 6.3 &     -2.2$\pm$0.2  & 3.7 \\ 
NGC2903 & -1.5$\pm$0.2 & 2.5 & -3.1$\pm$0.1 & 4.9 &     -2.4$\pm$0.2  & 3.7 \\ 
NGC3184 & -1.3$\pm$0.2 & 2.2 & -1.3$\pm$0.4 & 4.9 &     -1.8$\pm$0.4  & 4.2 \\ 
NGC3198 & -0.4$\pm$0.3 & 4.0 & -2.2$\pm$0.2 & 5.5 &     -1.8$\pm$0.3  & 5.9 \\ 
NGC3621 & -0.8$\pm$0.2 & 3.5 & -2.4$\pm$0.3 & 5.2 &     -2.0$\pm$0.1  & 4.9 \\ 
NGC4736 & -0.3$\pm$0.2 & 2.4 & -2.7$\pm$0.1 & 4.0 &     -2.4$\pm$0.3  & 4.5 \\ 
NGC5194 & -1.7$\pm$0.2 & 2.8 & -2.6$\pm$0.3 & 4.9 &     -2.1$\pm$0.2  & 2.8 \\ 
NGC5236 & -1.9$\pm$0.2 & 1.9 & -2.6$\pm$0.1 & 3.1 &     -2.0$\pm$0.1  & 2.4 \\ 
NGC5457 & -2.2$\pm$0.1 & 2.7 & -2.5$\pm$0.1 & 3.5 &     -2.3$\pm$0.1  & 5.6 \\ 
NGC6946 & -1.6$\pm$0.1 & 3.9 & -2.0$\pm$0.1 & 6.3 &     -1.6$\pm$0.2  & 3.9 \\ 
NGC7793 & -1.7$\pm$0.2 & 1.9 & -1.7$\pm$0.2 & 4.2 &     -2.0$\pm$0.1  & 2.0 \\ 
NGC2841 & -1.7$\pm$0.2 & 3.2 & -2.9$\pm$0.2 & 3.2 &     -1.8$\pm$0.2  & 5.9 \\ 
NGC3031 & -0.7$\pm$0.1 & 4.5 & -0.8$\pm$0.2 & 4.5 &     -0.7$\pm$0.1  & 4.5 \\ 
NGC3521 & -1.8$\pm$0.1 & 4.1 & -3.4$\pm$0.1 & 7.1 &     -2.6$\pm$0.1  & 5.7 \\ 
NGC5055 & -1.6$\pm$0.1 & 3.2 & -2.5$\pm$0.1 & 5.5 &     -2.2$\pm$0.1  & 3.2 \\ 
\hline
\end{tabular}
\end{center}
\label{tab:alpha}
\caption{Table giving the best fit power law index and the geometric mean of the range of baselines for the fit to the power spectra estimated from the visibility based and image based estimators. Image based power spectra are estimated from  both natural and robust weighted images. The baselines are in units of $k\lambda$.}
\end{table}
\begin{figure}[t!]

\subfloat[]{\includegraphics[scale=.35]{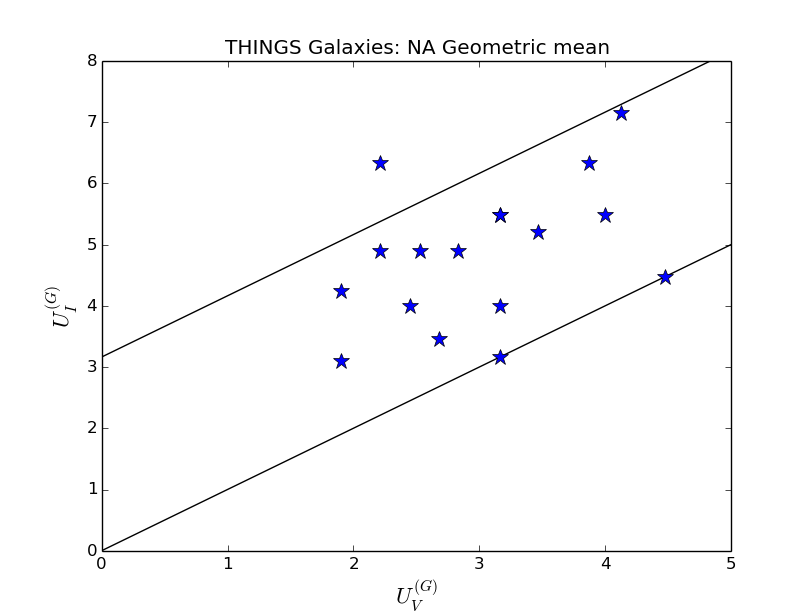}}
\subfloat[]{\includegraphics[scale=.35]{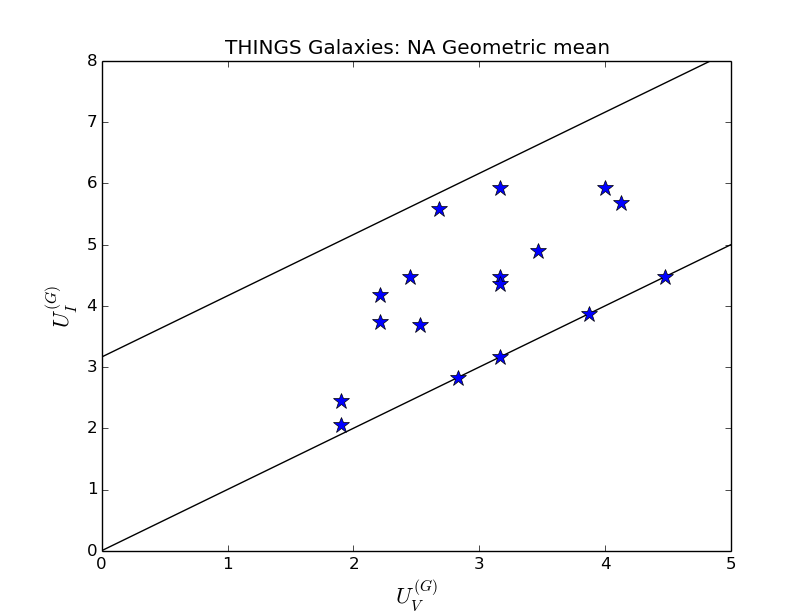}}
\caption{Figures showing the scatter plot of the geometric mean of the range of baselines for the power law fit with image and visibility based methods with natural weighted [Fig~(a)] and robust weighted [Fig~(b)] images.}\label{fig:ps2}
\end{figure}

We plot the values of $\alpha_{v}$ and $\alpha_{I}$ along with the error bars in figure~\ref{fig:ps1}. Left panel (a) corresponds to the robust weighted while the right panel (b) corresponds to the natural weighted map. The dashed line in each case  gives $\alpha_{v} = \alpha_{I}$. As it is clear for majority of the  galaxies we have studied here, the data points lie away from the equality line for both cases of natural and robust weighted maps. In fact, the image based estimator systematically produce a steeper spectra. This is exactly the same trend we have seen from the simulated observation. Though for three galaxies from the natural weighted image and five from the robust weighted image have $\alpha_{I} \sim \alpha_{V}$, on an average the two estimator gives statistically different results. Both the Pearson and Spearman correlation coefficients are also indicative of the lack of correlation between the two estimates of $\alpha$.

To check if we are not using completely different range of baselines to perform the power law fit for visibility and image based estimators, we also do a comparison plot of the geometric mean of the maximum and minimum baseline values of the two fits in figure~\ref{fig:ps2}. Note that, for the reasons discussed before, we do not expect the range of baselines the fits are performed in two different cases would be exactly same. The solid black line corresponds to a complete match. A typical range of baselines for which the power law fit  is performed using the visibility based estimator is $\sim 1 - 10 \ k\lambda$. The dashed line corresponds to $y = x+\sqrt{10}$. We see most of the galaxies lie within these two lines ensuring that we do perform the power law fit in similar range of baselines.

\section{Invariance in mean quantity}
\begin{figure}[t!]
\begin{center}

\includegraphics[scale=.45]{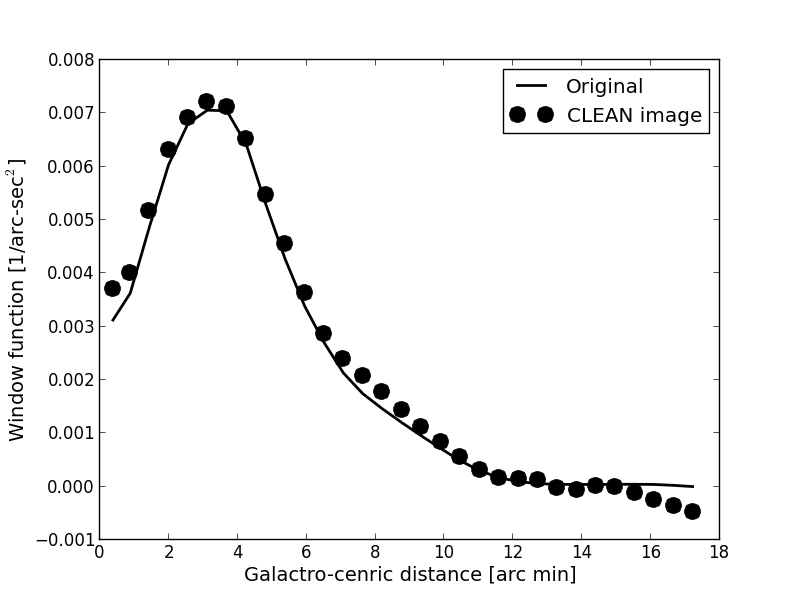}
\caption{Azimuthally averaged window function of the model galaxy (black solid line) and the reconstructed CLEANed image from simulation (black circles) are shown.}\label{fig:wind}
\end{center}

\end{figure}

Power spectrum is a second order statistical measure of the data. Various first order measures, for example local mean of the reconstructed images are of scientific interest. Azimuthally averaged window function is often used to access the radial variation of the \HI column density of the galaxy and compare it with stellar population, star formation rate etc. Unlike the power spectrum, we can not have a direct estimate of the window function from the observed visibilities, image reconstruction is necessary. We  estimate azimuthally averaged window function from CLEANed images from the simulated visibilities. This is shown with the discrete circles in  figure~\ref{fig:wind}.  The black solid line corresponds to the azimuthally averaged window function of the model galaxy. Clearly, the window function estimated from the power spectrum almost exactly follow the window function of the model. This demonstrates that the mean quantities can be estimated by reconstructed images from the visibility as the mean of the residual deconvolution noise is practically small.

\section{Discussion and Conclusion}
In this chapter we  perform controlled tests,  using simulated sky observations, that 
 quantify the reliability of the sky  brightness fluctuation
  power spectrum estimators. We see that the visibility based method reproduce the reference power spectrum from the input image. The image based estimator, however, is seen to have noise bias and the estimated power spectrum does not reproduce the reference spectra in any of the simulations. We expected that the power spectrum of the  dirty image would have a noise bias arising due to the mixing of scales by the interferometer beam. It is seen that even  for the deconvolved  image, obtained  using CLEAN with sufficiently low gain, the image based power spectra has sufficiently high noise bias.

We understand the mismatch between the image and visibility based power spectra in the following way. Interferometers sample the visibility function at discrete points in the baseline space. As the visibilities can be approximated as Fourier transform of the sky image, each measure of visibilities contains information of the entire observed field of view. In the case when the observed sky contain only a single point source, in principle, measuring the visibility at a single point in the baseline space would recover all the information of the source. Hence, in such a special case the sky image can be accurately reconstructed from the observed visibilities. As the number of point sources increase in the sky, it become more and more non trivial to infer the sky image from  the direct Fourier transform of the discretely sampled visibilities, the dirty beam. Reconstruction of the sky image requires to estimate comparatively rage number of parameters from the visibilities that represent the sky and a  deconvolution of the complicated dirty beam becomes necessary. Strictly speaking, for a sufficiently complicated sky image, with  incomplete baseline coverage, it is not possible to accurately reconstruct the sky. One well used procedure is CLEAN, which essentially assume the sky to be a collection of point source and try to estimate the position and the amplitude of these sources. It starts with identifying the brightest point source(s). However, as in the dirty image each of the point sources are scaled in amplitude and shifted by the side lobes of the other sources, such an estimation is always errors. Such an error is more apparent, when there are two sources very near to each other. To address this limitation, one use a loss factor ( usually called gain ). An extreme example is an extended structure, which CLEAN consider as many point sources next to each other. In such a case, using a sufficiently low loss factor also does not seem to reproduce the sky fairly accurately. An errors reproduction of the sky and hence fluctuations in the sky image, in tern gives an errors  power spectrum.

In order to estimate the power spectra, however, one do not need to know all the properties of the sky. Power spectra gives a statistical description of the sky and is assumed to be a smooth function. Hence, it need to be sampled at a relatively larger baseline values. Moreover, here we are estimating the isotropic power spectra. In fact, in the absence of the window function the power spectra can be modeled by a few parameters only. This makes the visibility based power spectrum estimator practically accurate. For any given baseline coverage, the visibility based estimator estimates the power spectra at the baselines where it is measured. Hence, though it spans only a subspace of the power spectrum function, at the given subspace the measurements are accurate. This is essentially reflected in our test. Hence, it is fair statement that the visibility based estimator estimates the power spectrum of the sky brightness fluctuations irrespective of the details of the baseline coverage. For an image based estimator the estimated power spectrum depend on the baseline coverage and the details of the deconvolution algorithms used to reproduce the image.

In our simulations we have not included any measurement noise to the visibilities ( as well as any uncalibrated gain variations, which also behave as noise ).  Inclusion of measurement noise would give rise to an additional correlated noise problem as discussed in Dutta (2013). However, these effects would depend on signal to noise of the measurement and can be made small enough (in principle) by  increasing the integration time of observations.

Here we also estimate the power spectra of the THINGS galaxies using the image based estimator starting from the CLEANed images form the THINGS data product. Considering the visibility based power spectra as reference we see that statistically the imaged based power spectra of THINGS also have significant noise bias, particularly,  the bias systematically makes the power spectra more steeper. In a few cases the image based spectra seem to agree with the reference spectra, however, it is clear from this investigation that image based power spectra should not be used without any characterization of the bias caused by the deconvolution noise.

Finally we note that though we use intensity of the \HI emission from nearby galaxies as models of the sky image, 
the limitations of he image based power spectra discussed here would be as relevant for any extended structure. We make a strong statement here that the results discussed in this work amounts in to rethink about  the use of the image based power spectrum in literature to infer various astrophysical phenomena. However, estimator of the  the mean quantities, like the azimuthally averaged window function from the reconstructed image are unbiased and can be used directly.

\newpage

\chapter{Visibility moment and the power spectrum of turbulent velocity} \label{ch: introduction}
In addition to the column density distribution, observing spectral lines we can also estimate the dynamical informations of the \HI gas in the galaxy. This is because the observed spectral lines are Doppler shifted  proportional to the line of sight velocities of the gas in the galaxy. However, we should note here that because velocity is a vector, unlike the column density, velocity correlation is given in terms of a matrix of dimension $3 \times 3$, where each element measures the correlations of one component of the velocity with other. Since observationally we can only probe the line of sight component of the velocity, we can estimate only one component of the velocity correlation matrix, that is line of sight to itself. 

Unlike the power spectrum of the column density, estimating velocity power spectrum from the radio interferometric observations is not straight forward. One way is to reconstruct the sky brightness distribution at different frequencies and  use that with various techniques like the Velocity Channel Analysis (VCA), the Velocity Coordinate Spectrum (VCS), the statistics of centroid of velocities to estimate the parameters of the velocity power spectrum. These methods were originally developed to measure the velocity power spectrum at relatively smaller scales ($\sim 1 - 100 $ pc), where mostly single dish observations were used.\footnote{It is possible to use visibilities directly for VCA.}  An inherent limitation of the reconstructed image from the interferometric data is discussed and investigated in detail in chapter 2, where we clearly demonstrate that the power spectra calculated from the reconstructed image are biased. Moreover, \citet{2015MNRAS.452..803D} has demonstrated that the VCS techniques have limitations when applied to external spiral galaxies. \citet{2016MNRAS.456L.117D} has introduced a velocity based  estimator for the turbulent velocity fluctuations. In this chapter we discuss this estimator, its implementation and application to a spiral galaxy NGC~6946.

\section{Visibility moment estimator}\footnote{The calculations presented in this section are reproduced from Dutta et al. (2016)}
For 21-cm observation of external spiral galaxies with existing radio telescopes like VLA,\footnote{Very Large Array, New Mexcico} GMRT,\footnote{Giant Meterwave Radio Telescope, Pune} the observed quantity is the visibility $\mathcal{V}(\vec{U},v)$, the Fourier transform of the sky brightness distribution:
\begin{equation}
\label{eq:V_u}
\mathcal{V}(\vec{U},v) = \int d\vec{\theta} \hspace{2.5pt}e^{i2 \pi \vec{U}.\vec{\theta}} I(\vec{\theta},v) 
\end{equation}
We define the zeroth moment of the visibilty as,
\begin{equation}
 V_0(\vec{U}) = \int dv\hspace{2.5pt}\mathcal{V}(\vec{U},v)
\end{equation}
 Using equation~\ref{eq:I_theta} and ~\ref{eq:phi}, we clearly see
\begin{equation}
V_0(\vec{U}) = \tilde{M}_0(\vec{U}),
\end{equation}
where $\tilde{}$ denotes Fourier transform. Similarly, we define the first moment of visibility as  
\begin{equation}\label{eq:V_u}
 V_1(\vec{U}) = \int dv\hspace{2.5pt}v\hspace{2.5pt}\mathcal{V}(\vec{U},v)
\end{equation}
Here the velocity integrals are over the entire spectral range of \HI emision. Using equation~\ref{eq:V_u} and~\ref{eq:I_theta}, we have 
  \begin{equation}
 V_1(\vec{U}) = \int d\theta\hspace{2.5pt}e^{i2 \pi \vec{U}.\vec{\theta}}\hspace{2.5pt}I_0\int dz \hspace{2.5pt}n_{HI}(\vec{r})v_z(\vec{r})
\end{equation}
where $v_z(\vec{r})$, line of sight velocity, has two components: (a) $v_z^{\Omega}(\vec{r})$, the line of sight component of the systematic rotation velocity of the galaxy and $v_z^T(\vec{r})$, the line of sight component of the random motion because of turbulence. For almost face on spiral galaxies with small inclination angle, systematic rotation can be assumed to be independent of z and thus we can   write 
\begin{equation}
v_z(\vec{r})=v_z^{T}(\vec{r})+v_z^{\Omega}(\vec{\theta})
\end{equation}
Hence we get,
 \begin{eqnarray}
\nonumber
 V_1(\vec{U}) &=&  \int d\theta\hspace{2.5pt}e^{i 2 \pi \vec{U}.\vec{\theta}}\hspace{2.5pt}M_0(\vec{\theta})v_z^{\Omega}(\vec{\theta})+
\int d\theta\hspace{2.5pt}e^{i2 \pi \vec{U}.\vec{\theta}}\hspace{2.5pt}I_0\int dz \hspace{2.5pt}n_{H1}(\vec{r})v^T_z(\vec{r}) \\ 
&=& V_0(\vec{U})\otimes\tilde{v}^{\Omega}_z(\vec{U}) +  \tilde{\chi}(\vec{U}),
\end{eqnarray}
where 
\begin{equation}
\chi(\vec{\theta})= I_0\int dz \hspace{2.5pt}n_{H1}(\vec{r})v^T_z(\vec{r})
\end{equation} 
 which contains information about both column density and turbulent velocity. 

Under the assumption that the random turbulent velocity and systematic rotational velocity of the galaxy are uncorrelated, the quantity $P_{\chi}(\vec{U})=\left<|\tilde{\chi}(\vec{U})|^2\right>$ is given by
\begin{equation}
P_{\chi}(\vec{U})= \left<|V_1(\vec{U})|^2- \tilde{C} (\vec{U})\right>
\end{equation} 
with
\begin{equation}
\tilde{C}(\vec{U}) = P_0(\vec{U})\otimes|v^{\Omega}_z(\vec{U})|^2
\end{equation}
Here $P_0(\vec{U})$ is the column density power spectrum of the galaxy. The angular bracket  denotes average over statistical ensembles. In practice, we are always provided with a single sky realization.  While calculating the column density power spectrum in chapter 2, we have bypassed this problem by performing an azimuthal average with the assumption of statistical homogeneity and isotropy. At the present case, neither $V_1(\vec{U})$ nor $\tilde{C}(\vec{U})$ are statistically anisotropic, and individual azimuthal averages to calculate them does not work. However, the complete right hand side (RHS) of above equation (without the angular bracket) is statistically isotropic. Hence, the ensemble average in the RHS of above equation can be replaced with an annular average when the average is done over the entire RHS expression. 

Here, $|V_1(\vec{U})|^2 $ and $P_0(\vec{U}) $ can be estimated directly from measured visibilities. Evaluation of the quantity $\tilde{C}(\vec{U})$ requires estimation of the line of sight component of the systematic rotational velocity of the galaxy. We shall discuss how this is performed in detail shortly. The quantity $P_{\chi}(\vec{U})$ has information on both power spectrum of column density and turbulent velocity. In case of the  spiral galaxies with thin disk, for  power spectrum estimated at very larger scales compared to the scale height of the galaxy 
\begin{equation}
P_{\chi}(\vec{U})=P_0(\vec{U})\otimes P^T_{v}(\vec{U})
\end{equation}
The power spectrum of the column density $P_0(\vec{U})$ can be estimated through the visibility moment zero. Estimating  the power spectrum of the turbulent velocity $P^T_{v}(\vec{U})=\left< |\tilde{v}^T_z(U)|^2 \right>$ then reduces to a problem of either model dependent regression analysis or direct deconvolution.
We shall discuss implementation of this estimator in the next section.
\section{Implementation of the visibility moment estimator}
In this section we describe in detail how the visibility moment estimator for $P_{\chi}(\vec{U})$ is implemented. First we  measure the quantity $v_z^{\Omega}(\vec{\theta})$ using shapelet reconstruction from the moment 1 map of the reconstructed specific intensity distribution. We use this to   estimate $\tilde{C}(\vec{U})$ and then $P_{\chi}(\vec{U})$. Finally we perform a parametric estimate of $P^T_{v}(\vec{U})$. 
 
\subsection{The systematic rotation  velocity}
We use the natural weighted moment one maps generated from the reconstructed specific intensities of the galaxy  to estimate $v_z^{\Omega}(\vec{\theta})$. Using the definition of the moment one map and eqn.~(3.6), we get for a thin disk galaxy with low inclination angle,
\begin{eqnarray}
M_1(\vec{\theta}) &=& \frac{\int dz\hspace{2.5pt}I_0n_{H1}(\vec{r})\left[v^T_z(\vec{r})+ v^{\Omega}_z(\vec{\theta})\right]}{M_0(\vec{\theta})} \\ \nonumber
&=& v^{\Omega}_z(\vec{\theta}) + \frac{\chi(\vec{\theta})}{M_{0}(\vec{\theta})}.
\end{eqnarray}
If we assume the density and velocity induced by turbulence is statistically homogeneous and isotropic and the moment zero map of the galaxy is azimuthally symmetric at large scales, average of the second term in the above expression is zero.
 Hence,
\begin{equation}
\nonumber
\langle {M}_1(\vec{\theta}) \rangle = v_z^{\Omega}(\vec{\theta}).
\end{equation}
The above average is in principle over an ensemble. Owing to the fact that we have only one sample of the sky, we need to choose an alternate averaging procedure. As the line of sight component of the rotational velocity vary at large scales compared to the scales where the turbulence is effective, we may choose to perform a local average of the moment one map. Here we need to determine the choise of the averaging scale and the  averaging kernel. An obvious choice is a Gaussian kernel with a length scale that is larger than the scales we would like to measure the velocity power spectrum, but smaller than the scales at which the rotational velocity may vary. However, a Gaussian kernel is azimuthally symmetric, while the rotational velocity at least have a dipole asymmetry. It would be good to rather perform a shapelet construction of the moment one map and choose the scale and order of the shapelet such that criteria for the averaging scale is satisfied.

\subsection{Construction of $\tilde{C}(\vec{U})$}
Line of sight component of the rotation velocity is a smooth function of the sky angular coordinates $\vec{\theta}$. We have discussed the method of estimating this in the previous section, where we get it in a grid in the $\vec{\theta}$ space. Hence the quantity $\mid \tilde{v}_{z}^{\Omega}(\vec{U}) \mid^{2}$ can be calculated by finding the modulus square of the discrete Fourier transform of $v_z^{\Omega}(\vec{\theta})$. We perform the convolution of $P_{0}(U)$ and $\mid \tilde{v}_z^{\Omega}(\vec{\theta}) \mid^{2}$ by multiplying their inverse Fourier transforms in a grid in the  $\vec{\theta}$ plane and applying the convolution theorem.

\subsection{Estimation of $P_{\chi}(\vec{U})$}
In order to find $P_{\chi} $, we need to find $|V_1(\vec{U})|^2$ and $\tilde{C}(\vec{U})$. It is straight forward to estimate $V_1(\vec{U})$  from measured visibility $\mathcal{V}(\vec{U},v)$ through its definition. However, as the measured visibilities are not in uniform grids in the baseline space, the values of  $V_1(\vec{U})$ are also not in grid. It is important to note here that  the recorded visibilities (and hence also $V_1(\vec{U})$),  contain sky signal and  noise parts:
\begin{equation}
V_1(\vec{U}) = S_1(\vec{U})+ \mathcal{N}
\end{equation}
In the interferometric observations, the individual measurements of $V_1(\vec{U})$  often does not have signal to noise higher than one. Hence, if we directly calculate $\mid V_1(\vec{U}) \mid^{2}$, it will be biased by the standard deviation of the noise. \citet{2009MNRAS.398..887D} has discussed how correlating the visibilities at nearby baselines reduces the noise bias. Here we consider an alternate approach. Since the quantity $\tilde{C}(\vec{U})$ is calculated in grids, we estimate  $\mid V_1(\vec{U}) \mid^{2}$ in grids by the following procedure.
We calculate the gridded values of  $V_1$, i.e,  $V_{1}(\vec{U}_{g})$ as,
\begin{equation}
V_{1}(\vec{U}_{g})=\sum_{i=1}^{N_{g}}V_1(\vec{U_i})
\end{equation}
where $N_{g}$ is the number of measured values in a particular grid $\vec{U}_{g}$.
Similarly we define
\begin{equation}
S_{1}(\vec{U}_{g})=\sum_{i=1}^{N_{g}} \mid V_1(\vec{U_i})\mid ^2.
\end{equation}
The measurement noise $\mathcal{N}$ is uncorrelated across different baselines. In this case, the following combination of $V_{1}(\vec{U}_{g})$ and $S_{1}(\vec{U}_{g})$ gives an unbiased estimates of $\mid V_{1} \mid^{2}$ at the grid points $\vec{U_{g}}$, which we denote as $\mid V_{1} \mid^{2} (\vec{U} _{g})$,
\begin{equation}
|V_{1}|^2(\vec{U}_{g}) =\frac{\left[ (V_{1}(\vec{U}_{g})^2-S_{1}(\vec{U}_{g}) \right]}{^{N_{g}}C_2}
\end{equation}
where $|V^G_1(\vec{U})|^2$ are estimated in grids. Since $\tilde{C}(\vec{U})$ is a smooth function of $\vec{U}$ and are already estimated in a grid in $\vec{U}$, we can interpolate it to the baselines $\vec{U}_{g}$ . Finally, we use annular average over bins of $\vec{U}$ to find $P_{\chi}(U)$.
\subsection{Parametric deconvolution of the velocity power spectrum}
As mentioned in the beginning, $P_{\chi}( U )$ has information on both power spectrum of column density and turbulent velocity.
We have been considering the  particular case of spiral galaxies with thin disk, where 
\begin{equation*}
P_{\chi}(\vec{U})=P_0(\vec{U})\otimes P^T_{v}(\vec{U}).
\end{equation*}
Method to estimate the quantities $P_{\chi}(\vec{U})$ and $P_{\chi}(\vec{U})$ have been discussed already. Hence estimating the power spectrum is to perform a two dimensional deconvolution. Though a direct deconvolution may be possible, we prefer to perform a parametric estimate here. Since the power spectrum of the turbulent velocity is expected to be a power law, we assume a two parameter model for $P^T_{v}(\vec{U})$, given by
\begin{equation}
P^T_{M}(\vec{U})=A_{v}\hspace{2.5pt}U^{\beta}.
\end{equation}
We estimate the two parameters of $ P^T_{v}(\vec{U})$
by regression analysis. However, we do not follow the most common chi-square minimization of the parameters here. Usual chi-square method assume that the models are exact (that is does not have associated error) and weight the residual by the error in the measurements. Here, since part of the model ($P_0(\vec{U})$) is  estimated from the data, the model values also have uncertainties. We perform a monte-carlo based regression analysis, where we use several realizations of the model and the measured values and estimate the best fit parameters for each of these realizations. The statistics of the best fit values over the realizations are used then to access the parameters and their errors. For each set of values of the parameters we calculate 
\begin{equation}
P_{\chi}^M(\vec{U})=P_0(\vec{U})\otimes P^T_{M}(\vec{U})
\end{equation}
at the same grid as that of $P_{\chi}(\vec{U})$ and compare it with the measured value of $P_{\chi}(\vec{U})$ by estimating
\begin{equation}
  \Delta (A, \beta) = \frac{\left( P_{\chi}^M(\vec{U}_{G})-P_{\chi}(\vec{U}_{G}) \right)^2}{N_{G}^2}
\end{equation}
where $N_{G}$ is the number of grid points. We minimize the $\Delta$ with respect to $A$ and $\beta$ and find the best fit value. For those best fitted parametric values, we can find $P^T_{v}(\vec{U}) $.

\subsection{Monte-carlo error analysis}
In the previous sections we have described the steps to estimate the velocity fluctuation power spectrum from the radio interferometric observations. Any observational  estimates need to be supplemented by the statistical error associated with it. To estimate the errors  first note the errors in the directly measured quantities. In this case the measurement errors are associated with the errors in measuring the velocity and the visibilities. For the relatively straight forward statistical estimators, it is possible to find the errors in the estimated quantities by propagating the errors analytically from the measurements. However, for the visibility moment estimator we described above, analytical estimation is complicated. In such a case we may  do repeated observation of same sky. This is impractical given the limited observational time available at the telescope facilities. 

A solution to this problem is to produce synthetic observations that preserves the statistics of the measurements. For example, as the visibilities are known to have Gaussian random noise associated with their measurements, we use the mean and the standard deviation of the measured visibilities to generate different statistical realizations.  Considering  each of these realizations as an individual measurement, we can use the visibility moment estimator to get the parameters of the line of sight velocity fluctuation power spectrum. This gives us the probability distribution of the estimated quantities, their mean and standard deviation. The mean of the estimator hence calculated is taken to be the best estimate and the standard deviation is taken to be the error associated with it. This method of estimating the error is often termed as the monte-carlo estimation of the errors. 

Here we use a mixture of monte-carlo and analytical error-propagation methods to get to the errors in measured parameters of the velocity power spectrum.
In the previous chapter we have described the visibility based estimator for the column density power spectrum $P_0(\vec{U})$. The visibility based estimator also provides us with error estimates on the annular bins in baselines where the power spectra are measured. We assume a Gaussian  distribution to generate different realizations of the $P_0(\vec{U})$, that is
\begin{equation}
P^R_0(U) = P_0(U) + \sigma_{P_0}(U) \times \hat{n},
\end{equation}
where $\sigma_{P_0}(U) $ is the 1 $\sigma$ error is estimating $P_0(U)$ and $\hat{n}$ is a random number drawn from a normal distribution (Gaussian distribution with unit standard deviation and zero mean). The superscript $R$ stand for a particular realization.

The typical errors in the measured velocities arise from the width of the frequency channels of the observation. For the THINGS galaxies it is about  $5$ km sec$^{-1}$. We use the moment one map to estimate the line of sight component of the rotational velocities $v^{\Omega}(\vec{\theta})$. The moment one maps are generated using the reconstructed images from the visibilities, their local average provides us unbiased estimate of $v^{\Omega}(\vec{\theta})$ (section 2.2.1 and chapter 2). We use these estimates of $v^{\Omega}(\vec{\theta})$ and assume a Gaussian error of standard deviation of $5$ km sec$^{-1}$ to generate different realization of the $v^{\Omega}(\vec{\theta})$. These realizations combined with the different realizations of $P_{0}(U)$ discussed above are used to generate  different realizations of $\tilde{C}(\vec{U})$. We calculate the mean and standard deviations of these estimates and use them for further analysis. We estimate the errors in the estimates of $\mid V_{1}(\vec{U})\mid^{2}$ using similar analytical steps as that we use for $P_{0}(U)$. The final estimates of the errors in $P_{\chi}(\vec{U})$ is done by doing analytical error propagation from the errors in $\tilde{C}(\vec{U})$ and $\mid V_{1}(\vec{U})\mid^{2}$. 

\section{Data identification for application of the visibility moment estimator}
The visibility moment estimator for the power spectrum of the line of sight turbulent velocity of the external galaxies can be applied to the external spiral galaxies with low inclination angle. As the inclination angle increases the assumption that the rotation velocity is independent of the line of sight direction fails and the uncertainty in the estimates of the velocity power spectrum increases. \citet{2016MNRAS.456L.117D} have validated the estimator by using noise free simulated data. They showed that for low inclination angles, $i < 40^o$, this estimator does a better job with less uncertanity.  In their simulation they have chosen a simple model for line of sight velocity component with constant inclination angle, such that they avoid the uncertanity in the rotation curve estimation. In reality however, the galaxy's disk are not exactly flat, both inclination and position angle varies with radial distance. This can induce  significant uncertanitity in the estimation of rotation curve, for considerably lower inclination angle ($< 15^{o}$).  Hence we prefer to choose the inclination angle to lies in between $\sim 20^o$ to $\sim 35^o$. We also would like to avoid too much warp in the galaxy's disk, that is large variation in the inclination and position angles. The other important criteria to choose a galaxy is that it must have thin disk, adequate resolution in both angular and frequency axis.

We use radio interferometric data product from the THINGS \footnote{THINGS: The \HI Nearby Galaxy Survey (Walter et al. 2008)} survey, where they have observed 23 spiral galaxies. We choose the galaxy NGC~6946 which satisfy the above criteria. It has an inclination angle of $33^o$. The scale length of the galaxy is $\sim 20$ kpc whereas the upper limit of the scale height \citep{2013NewA...19...89D} is 300 pc. The angular resolution corresponds to the data is about $75$ pc, where as the velocity resolution is $5$ km sec $^{-1}$.
\begin{figure}[t!]
\subfloat[]{\includegraphics[scale=.39]{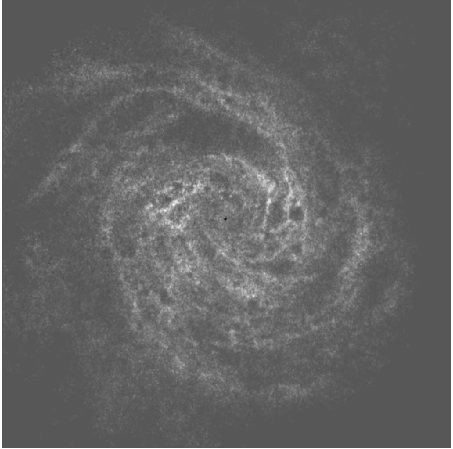}}
\hspace{10pt}
\subfloat[]{\includegraphics[scale=.39]{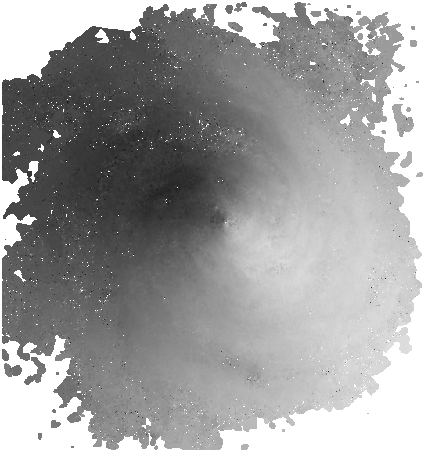}}
\caption{Fig~(a) is the moment zero map of NGC6946 and Fig~(b) is the moment one map of the same.}
\label{fig:NGC6946}

\end{figure}
In our analysis we would need both the visibility data from \HI emission and the moment one maps of the reconstructed specific intensity distribution. Along with H1 21 cm line emision, the direct observed visibilities originally also have continuum emission (mainly synchrotron emission).  We have modelled the continuum emission  from the line free frequency channels. This has been removed by UVSUB task in AIPS. We use the continuum subtracted visibilities for further analysis. THINGS \HI survey also provide moment maps of H1 emision for different external galaxies, generated by multi scale cleaning of the visibility data. Fig~\ref{fig:NGC6946} shows the moment zero and moment one map of NGC~6946 galaxy. We have used this moment one map for estimation of $\tilde{C}(\vec{U})$.

\section{Velocity power spectrum of NGC~6946}
In this section, we discuss about the results of the implementation of the visibility moment estimator on NGC 6946 galaxy. Following sections discuss about each step in detail and results obtained.

\subsection{Shapelet construction of $v^{\Omega}_z(\vec{\theta})$}
\begin{figure}[t!]

\subfloat[]{\includegraphics[scale=.35]{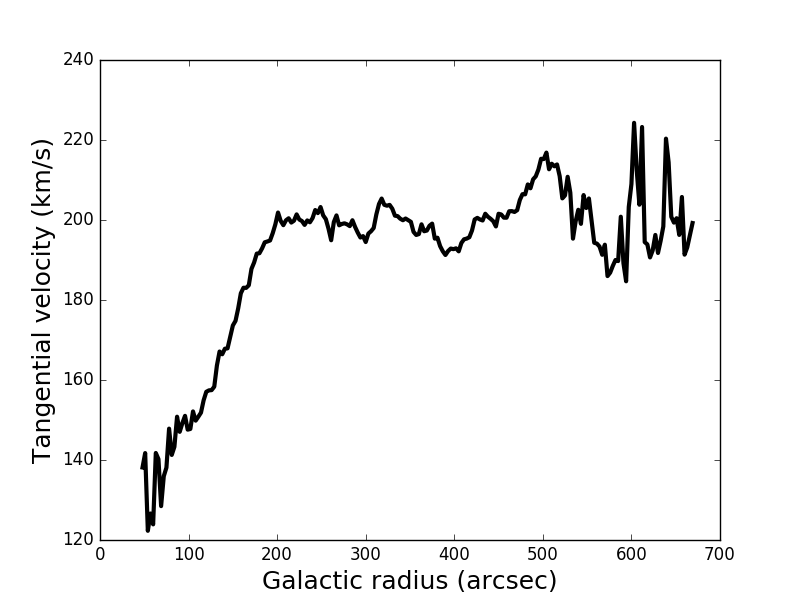}}
\hspace{10pt}
\subfloat[]{\includegraphics[scale=.35]{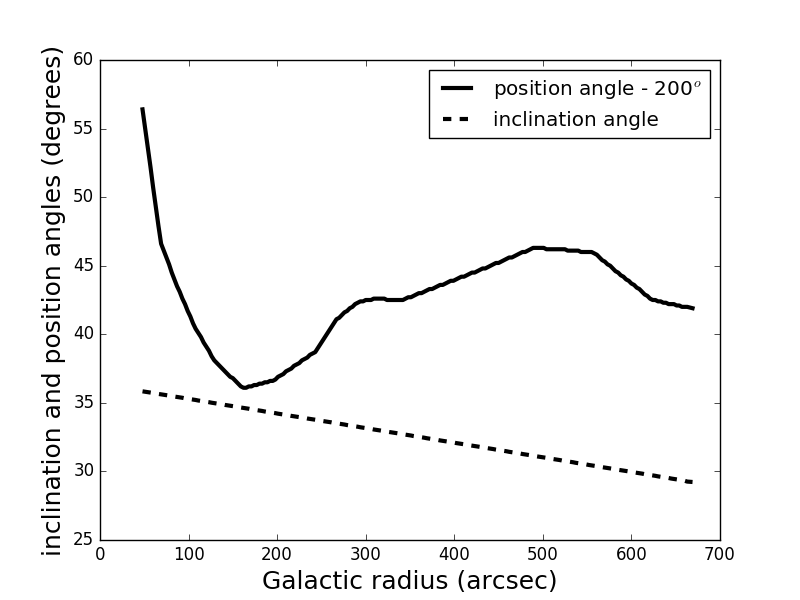}}
\caption{Fig~(a) is the rotation curve of NGC 6946 and Fig~(b) is the plot showing position angle and inclination angle as a function of galacto-centric radius. The position angle is offset by $200^{o}$ to show then in the same plot. Note that the inclination angle vary almost as a linear function of the galactocentric radius}\label{fig:rot_curve}
\end{figure}
We estimate, the line of sight component of rotational velocity $v^{\Omega}_z(\vec{\theta})$ using shapelet decomposition. In chapter 2, we have performed shapelet construction of the window function. We grossly follow the same procedure here with appropriate choice of the shapelet scale $\eta$  and the highest shapelet order  $n_{max}$. Both these parameters are necessary to identify the large scale component in the moment one map that  corresponds to the line of sight component to the rotation velocity. Later has contribution  from tangential rotational velocity, inclination angle and position angle of the galaxy. Using the reconstructed \HI image data cube, \citet{2008AJ....136.2648D} have estimated the rotation curve, position angle and inclination angle as a function of galactocentric radius. These are shown in figure~\ref{fig:rot_curve}. We estimate the 1D power spectrum of each of these three components to identify the large scale fluctuations of these curves.   These power spectrums, normalized to the standard deviation of corresponding curve are shown in figure~\ref{fig:PS_curve}.
 \begin{figure}[t!]
\begin{center}

\includegraphics[scale=.45]{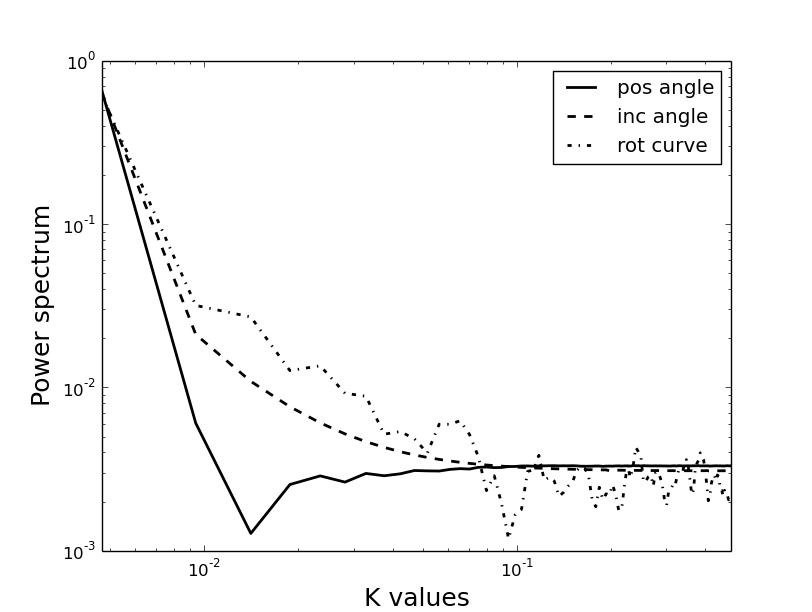}
\caption{One dimensional power spectrum of the tangential rotational velocity, inclination angle and position angle as a function of the inverse angular scale.}\label{fig:PS_curve}
\end{center}
\end{figure}
Clearly, the power spectra  has considerably more power at the larger scales. We considered the angular scale  where the all three power spectra goes below $1\%$ of peak value, as the scale for the shapelet construction. In this case, this happens at an angular scale of 157 arcsec, which corresponds to 7.5 kpc in the galaxy's disk. It is important to note that the estimates of the velocity power spectra we do here would be limited to 7.5 kpc at the larger scale side.
Next we do shapelet decomposition of moment one map of the galaxy with different $n_{max}$. For sufficiently large value of $n_{max}$, the shapelet construction would peak up the small scale fluctuations. We identify the highest value of $n_{max}$ for which in the reconstructed map fluctuations at scales smaller than $\eta$ are absent. In our case, $\n_{max} = 8$ came out to be this optimal value and we use it for further analysis.

   \begin{figure}[t!]
\subfloat[]{\includegraphics[scale=.43]{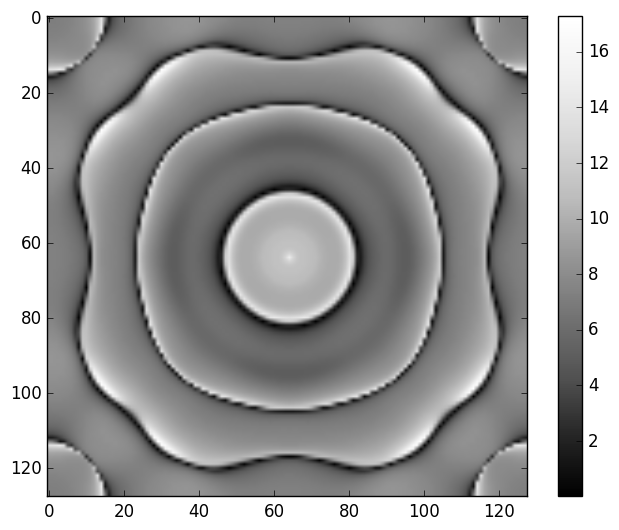}}
\hspace{10pt}
\subfloat[]{\includegraphics[scale=.60]{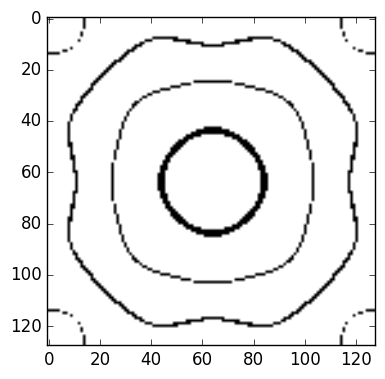}}
\caption{Fig~(a) show the SNR plot of the  $P_{\chi}(\vec{U})$ and Fig~(b) highlights (black) the regions with SNR $<5$.}\label{fig:SNR}
\end{figure}  
We generate 128 realizations of moment one map and use the above parameters to estimate $v^{\Omega}_z(\vec{\theta})$. We calculate  the mean and standard deviation of these maps. We use the column density power spectrum as given in Dutta et al. (2013) and the above realizations of the maps to generate 128 realizations of $C(\vec{\theta})$ and the maps corresponding to the mean and standard deviation of these maps. Figure~\ref{fig:SNR} (a) show the signal to noise ratio (snr) of $C(\vec{\theta})$, where Figure~\ref{fig:SNR} (b) shows in black the pixels in the snr map with snr$<$5. Clearly, most of the map has signal to noise more than 5 suggesting that the estimated $C(\vec{\theta})$ is statistically significant.
\begin{figure}[t!]
\begin{center}
\includegraphics[scale=.55]{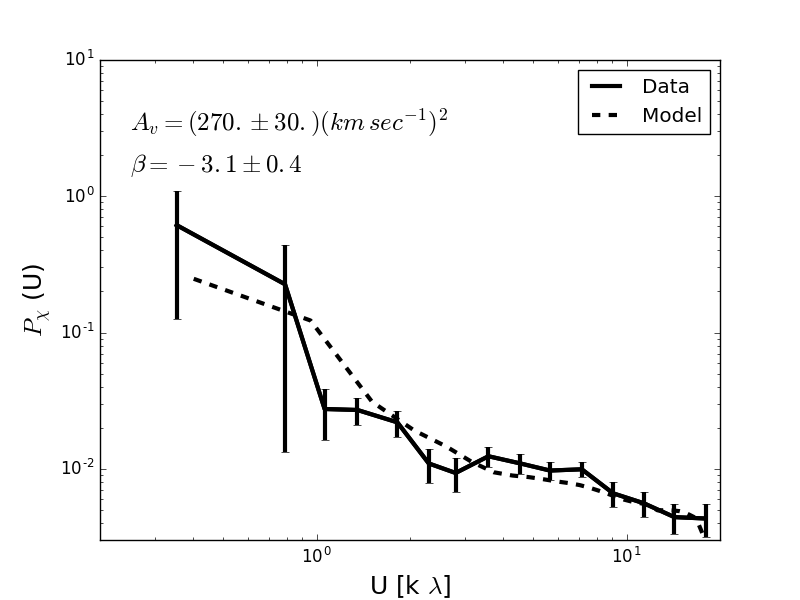}
\caption{Estimated (solid black) $P_{\chi}(\vec{U})$ of NGC 6946 galaxy and best fit model $P^M_{\chi}(\vec{U})$ (dashed black) for a power law model of velocity power spectrum.}\label{fig:PS_curve}
\end{center}

\end{figure} 
\subsection{Estimation of velocity power spectrum for NGC~6946}
We estimate $|V_1(\vec{U})|^2$ in a uniform grid in u-v plane from measured visibility data following the  procedure  discussed in section 3.2.3. Fourier transform of  $C(\vec{\theta})$ are estimated on the same u-v grids and subtracted from the $|V_1(\vec{U})|^2$ values. The resultant quantity is expected to be statistically homogeneous and isotropic. We estimate the $P_{\chi}(\vec{U})$, by doing annular average over these for the 128 realizations. The mean and standard deviation of the measured $P_{\chi}(\vec{U})$ of these realizations is plotted in  Figure~\ref{fig:PS_curve} with solid black line.

The estimated $P_{\chi}(\vec{U})$ can be used for the measurement of velocity power spectrum. For thin disk galaxy,  $P_{\chi}(\vec{U})$ is a two dimensional convolution of the column density map with the power spectrum of the line of sight component of the turbulent velocity. As mentioned before, we use a two parameter model for the visibility power spectrum and try to find the best fit values of the parameters. An obvious choice for the functional form of the velocity power spectrum is a power law,
\begin{equation*}
P^T_{M}(\vec{U})=A_{v}\hspace{2.5pt}U^{\beta}
\end{equation*}
For each realization of $P_{\chi}(\vec{U})$, we estimate the best fit parameter values for A and $\beta$ and calculate the mean and standard deviation over the realizations. Dashed line in the Figure~\ref{fig:PS_curve} shows the mean value of  best fit model $P^M_{\chi}(\vec{U})$. The best fit value for parameter, in the units of the power spectrum we use  are: $A_{v} =(270.\pm 30.)\times 10^6$ [km sec$^{-1}$]$^{2}$ and $\beta=-3.1\pm 0.4$. 

\section{Discussion and Conclusion}
We adopt a distance to the galaxy NGC~6946 to be $5.5$ Mpc (Walter et al. 2008). Dutta et al. (2013) have estimated the column density power spectrum of this galaxy, where they found that for baselines less than $1.5$ k$\lambda$ the window function effect dominates. At these baselines and lower, we also have relatively small number of independent realizations of the estimates of the power spectrum and hence sample variance dominates. We would like to restrict our interpretation within the baseline values $1.5$ k$\lambda$ to $10.$ k$\lambda$, that is the range over which the column density power spectrum was measured. These baselines corresponds to a length scale range of $300$ pc to $4$ kpc at the galaxy. Based on the amplitude of the velocity spectrum, at $4$ kpc we see velocity fluctuation of magnitude $9$ km sec$^{-1}$, whereas at $300$ pc it is as low as $1.2$ km sec$^{-1}$. The fluctuations at the largest scale is comparable to the typical vertical velocity dispersion in the spiral galaxies (Lewis 1984). Certainly at the smallest length scales the velocity fluctuations are comparable to thermal velocity fluctuations by the cold neutral medium. \citet{2009AJ....137.4424T} has estimated the velocity dispersion of the galaxy NGC~6946 ( amongst other galaxies), where they find the median velocity dispersion is about $10.1$ km sec$^{-1}$, consistent with the amplitude of the velocity fluctuations we measure here. They estimated the variation of the velocity dispersion as a function of radius and see that it reduces with increasing galactocentric radius. They compare the velocity dispersion with the star formation rate and discuss that within the radius of $8$ kpc, where all the star formation happens, feed back from the star formation is enough to maintain the observed dispersion. At larger radius, they conclude, other mechanisms like magneto rotational instability (MRI) etc may be driving the velocity dispersion. Our result suggest a large scale correlated velocity fluctuations at scales at least comparable to the star forming disk. \citet{2009ApJ...692..364F} investigated the effect of solenoidal and compressive forcing for compressible fluid turbulence. They found that the resultant velocity spectrum (in our definition) have a slope of $-2.89$ for solenoidal forcing, while for compressive forcing the slope is $-3.03$. Though our measurement of the slope $-3.1 \pm 0.4$ favours the compressive forcing slightly, it can not completely rule out the other. If we consider that the compressive forcing is favoured, MRI would not be an effective mechanism as the forcing associated would be more solenoidal than compressive in nature. A possible forcing may arise from the rotational instability caused by the density fluctuation in the dark matter halo that hosts the galaxy. We would like to investigate in this line in future.

\chapter{Discussion and Future Scope}
In this thesis we investigated the statistical properties of large scale density and velocity fluctuations from the radio interferometric observation of neutral hydrogen of external spiral galaxies. We compared different estimators of the column density fluctuation using numerical simulation. We implemented  the visibility moment estimator of the velocity power spectrum and estimated the power spectrum of the line of sight velocity fluctuations of the spiral galaxy NGC~6946. 

In first part of our work, we checked the efficacy of the visibility and the image based power  spectrum estimators using simulated observation of a model galaxy. We found that the image based power spectrum estimator have a noise bias which create a deviation from the true power spectrum. On the other hand, the visibility based power estimator recover the true power spectrum without any bias. The  reason for the deviation is the sparce baseline coverage and the  noise bias arising from the incomplete deconvolution is correlated with baseline distribution. Hence, the image based power spectrum estimator must be avoided for the radio interferometers with incomplete baseline coverage. It is important to note that it is not possible to estimate  the locally averaged quantities   directly from visibilities, image reconstruction is necessary. However, the mean of the deconvolution noise  in the image is zero and the locally averaged quantities we estimate from the reconstructed image do not have any significant bias. We demonstrated this by inspecting the  azimuthally average window function. We compared the  column density power spectrum of 18 spiral galaxies from the THINGS sample between image based and  visibility based power spectrum estimators. We found  that the power spectrum estimated from reconstructed images systematically follow a  steeper power slope. This suggests that the statistical study of the radio sky would require recording of the visibilities from the observations. For the future telescopes like the Square Kilometer Array (SKA) it is often argued that because of the huge data rate only images reconstructed in real time from the visibilities will be recorded. However, our result points out the importance of recording the visibilities.

We implemented the visibility moment estimator and measured the turbulent velocity power spectrum of the spiral galaxy NGC~6946. A power law model spectrum offer a statistically significant fit to the data suggesting a scale invariance velocity fluctuations at length scales as large as 4 kpc. Scale invariance density and velocity fluctuations for this galaxies at these scales suggest strongly for a turbulent dynamics. We found that the amplitude of the fluctuation at 4 kpc is about $9$ km sec$^{-1}$, consistent with the velocity dispersion of the \HI gas. The apparently larger errors associated with our measurement of the slope of the spectrum do not allow us to comment strongly on the nature of the driving mechanism, though it weakly favour compressible forcing at large scale. We guess that this can be the result of density fluctuations coupled by gravity compared to MRI driven fluctuations. {\bf Nevertheless, this thesis presents the first ever measurement of the  power spectrum of the line of sight velocity fluctuations of any external spiral galaxy (NGC~6946 here) and strongly indicate presence of large scale turbulence. }

The limitations in pointing out the nature of the driving force in this work arises mainly because of the measurement noise. This can be trivially improved by increasing both the observing time and baseline coverage. In practice, however, the THINGS survey already combine the B, C and D array of VLA and have significantly large integration time. The importance of the science objective here would surely drive future observations to improve on the data. Also, observations taken from other radio telescopes like GMRT, WSRT etc can be combined to the VLA data to improve upon the baseline coverage. In this work we used a parametric estimation of the velocity power spectrum. However, in principle it is possible to perform a direct deconvolution of the velocity spectrum from $P_{\chi}$, using well established algorithms like the Richarson-Lucy deconvolutions etc. Finally we note that we plan to estimate the velocity power spectrum of a sample of spiral galaxies (THINGS have four more galaxies in the right inclination angle window) in future, to access the true nature of the large scale fluctuations in more detail.

\bibliographystyle{ifacconf}
\bibliography{meera}
\addcontentsline{toc}{chapter}{Bibliography}

\end{document}